\documentclass[onecolumn]{article}
\usepackage[utf8]{inputenc}
\usepackage[a4paper, total ={15cm,22cm}]{geometry}
\usepackage{graphicx}
\usepackage[rightcaption]{sidecap}
\usepackage{authblk}

\graphicspath{ {./images/} }

\usepackage[style=ieee, citestyle=numeric-comp,sorting=none]{biblatex}
\usepackage{hyperref}
\hypersetup{
    colorlinks=true,
    citecolor=blue,
    linkcolor=black,
    urlcolor=blue
}
\usepackage{parskip}
\usepackage{enumerate}
\usepackage{enumitem}
\usepackage{algorithm2e}
\usepackage{amsmath,amsfonts,amssymb}
\usepackage{caption}
\usepackage{subcaption}
\usepackage[capitalize,noabbrev]{cleveref}
 \DeclareMathOperator*{\argmax}{arg\,max}
 \DeclareMathOperator{\sech}{sech}
\usepackage{cancel}
\usepackage[symbols,nogroupskip,nonumberlist,nopostdot=false,abbreviations]{glossaries-extra}
\usepackage{array}
\usepackage{booktabs}
\usepackage{colortbl}

\usepackage{soul}

\newtheorem{lemma}{Lemma}

\newtheorem{remark}{Remark}
\newabbreviation{ode}{ODE}{Ordinary Differential Equation}

\newglossaryentry{ess}{
    name={ESS},
    description={Evolutionary Stable Strategy},
    first={Evolutionary Stable Strategy (ESS)},
    plural={ESSs},
    firstplural={Evolutionary Stable Strategies (ESSs)},
}

\newabbreviation{dfe}{DFE}{disease-free equilibrium }
\newabbreviation{mfe}{MFE}{mutant-free equilibrium }
\newabbreviation{covid}{SARS-CoV-2}{severe acute respiratory syndrome coronavirus 2}
\newabbreviation{hcv}{HCV}{Hepatitis C virus}
\glsxtrnewsymbol[description={basic reproduction number}]{r0}{\ensuremath{\operatorname{\mathcal{R}_0}}}
\glsxtrnewsymbol[description={A birth function with respect to a population of size $N$ and exclusively contributes to the susceptible compartment }]{bN}{\ensuremath{b(N)}}
\glsxtrnewsymbol[description={Susceptible compartments}]{S}{\ensuremath{S}}
\glsxtrnewsymbol[description={disease compartments for strain $j$ for $j=r,\ m$}]{I}{\ensuremath{\mathbf{X}_j}}
\glsxtrnewsymbol[description={Recovered compartments}]{R}{\ensuremath{R}}
\glsxtrnewsymbol[description={A natural death function}]{dA}{\ensuremath{d(.)}}
\glsxtrnewsymbol[description={Transmission rates of the disease with strain $j$ for $j=r,\ m$}]{fj}{\ensuremath{f_j}}
\glsxtrnewsymbol[description={Recovery rates of the disease with strain $j$ for $j=r,\ m$}]{gj}{\ensuremath{g_j}}
\glsxtrnewsymbol[description={A diagonal matrix of susceptibles that is adjusted to calculate the transformation from \gls{S} to the infected compartments}]{AdjS}{\ensuremath{\mathbf{S}}}

\glsxtrnewsymbol[description={The transmission matrix (may depend on the state variables) whose elements correspond to transmission events in which an epidemiological infection (\gls{I}) is acquired \cite{diekmann2010construction} }]{Finj}{\ensuremath{\mathbf{S}\mathbf{F}_j}}

\glsxtrnewsymbol[description={A transition matrix whose elements corresponding to all other changes of the disease compartments \gls{I} \cite{diekmann2010construction}}]{Dinj}{\ensuremath{\mathbf{D}_j}}
\usepackage{tcolorbox}
\usepackage{siunitx}
\newenvironment{myframe}[1]{%
    \begin{tcolorbox}[colback=white,colframe=black,title=#1]
    }{\end{tcolorbox}
}
\newlist{myenumerate}{enumerate}{1}
\setlist[myenumerate]{label=\textbf{\arabic*.}}

\newlist{myitemize}{itemize}{1}
\setlist[myitemize]{label=\textbf{$\bullet$}}

\newlist{mydescription}{description}{1}
\setlist[mydescription]{font=\normalfont\bfseries}
\makeglossaries
\newcommand{\keywords}[1]{\def\@keywords{#1}}
\title{The context-specificity of virulence evolution revealed through evolutionary invasion analysis}

\author[1,2]{Sudam Surasinghe}
\author[1]{Ketty Kabengele}
\author[1,3]{Paul E. Turner} 
\author[*,1,2,4,5]{\\ C. Brandon Ogbunugafor}
\affil[1]{Department of Ecology and Evolutionary Biology, Yale University, New Haven, CT, 06520 USA}
\affil[2]{Public Health Modeling Unit, Yale School of Public Health, New Haven, CT 06510 USA}
\affil[3]{Microbiology Program, Yale School of Medicine, New Haven, CT 06510 USA}
\affil[4]{Santa Fe Institute, Santa Fe, NM, 87501 USA}
\affil[5]{Vermont Complex Systems Center, University of Vermont, Burlington, VT, 05405 USA}
\affil[*]{For correspondence:  \texttt{brandon.ogbunu@yale.edu}}
\date{}

\begin{document}
\maketitle
\begin{abstract}
\noindent
Models are often employed to integrate knowledge about epidemics across scales and simulate disease dynamics. While these approaches have played a central role in studying the mechanics underlying epidemics, we lack ways to reliably predict how the relationship between virulence (the harm to hosts caused by an infection) and transmission will evolve in certain virus-host contexts. In this study, we invoke evolutionary invasion analysis---a method used to identify the evolution of uninvadable strategies in dynamical systems---to examine how the virulence-transmission dichotomy can evolve in models of virus infections defined by different natural histories. We reveal that peculiar ecologies drive different evolved relationships between virulence and transmission. Specifically, we discover patterns of virulence evolution between epidemics of various kinds (SARS-CoV-2 and hepatitis C virus) and that varying definitions of virulence alter our predictions for how viruses will evolve. We discuss the findings in light of contemporary conversations in the public health sector around the possibility of predicting virus evolution and in more extensive theoretical discussions involving virulence evolution in emerging infectious diseases. 
\end{abstract}
\section*{Introduction}
Recent events have reinvigorated interest in the evolution and ecology of infectious disease, specifically, what rules (if any) govern how lethal a given pathogen will become in a population of hosts. These questions have formed a theoretical canon defined by hundreds of studies and analytical descriptions of the evolvability and constraints surrounding how a pathogen evolves increased virulence \cite{anderson_coevolution_1982,ewald_host-parasite_1983,ewald_evolution_2004,bull_virulence_1994,lenski_evolution_1994,frank_models_1996, ebert_optimal_1997}. Virulence can be defined in many ways, but mainly relates to some measure of harm done to hosts by pathogens or the capability of causing disease in host organisms \cite{read_evolution_1994,casadevall_hostpathogen_2001,thomas_pathogenicity_2004}. These ideas have been applied to pathogens of various kinds—parasitic, helminthic, bacterial, and viral—infecting a vast number of host types, from plants to nonhuman animals, and humans \cite{frank_models_1996,alizon_multiple_2013,cressler_adaptive_2016}. Classically, it is framed in terms of its relationship to transmission, applying to a suite of traits contributing to a pathogen’s ability to successfully transmit an infection from one host to another \cite{lipsitch_virulence_1997,bull_theory_2014}. 



One of the goals of the evolution of virulence canon is to predict how virulence will change in an evolving interaction between pathogen and host. This is especially relevant in the context of viral pathogens (especially RNA viruses), where the rapid evolution of viruses renders the ecological and evolutionary scales similar \cite{steinhauer_rapid_1987,pybus2009evolutionary,duffy_why_2018}. These ideas rose to prominence during the COVID-19 pandemic, in which an array of opinions arose regarding how SARS-CoV-2 populations would evolve with respect to their virulence \cite{grubaugh_we_2020,kissler_projecting_2020,van_dorp_no_2020,alizon_sarscov2_2021}. 


Mathematical modeling was foundational in the historical development of epidemiology \cite{brauer_mathematical_2017,siettos_mathematical_2013, kucharski2020rules,jones2020shape} and has continued to serve a critical role in the study of infectious outbreaks \cite{lofgren_mathematical_2014,cobey_modeling_2020}, providing insights for clinical interventions and public health policies \cite{whitty_infectious_2014,heesterbeek_modeling_2015}. In addition, models can serve as instruments to explore theoretical questions or to examine how to predict the dynamics of epidemics \cite{scarpino_predictability_2019}. Some methods have been pioneered to identify how a given pathogen will evolve under a given set of circumstances. These methods are related to the notion of the evolutionary stable strategy (ESS), first pioneered in the study of evolutionary game theory, which describes an optimal, “uninvadable” strategy \cite{smith1973logic, otto2011biologist, smith1982evolution,vincent_evolutionary_2005,bukkuri_evolutionary_2021}. This perspective has since been applied broadly in infectious contexts, including virus evolution in the setting of different multiplicity of infections \cite{turner1999prisoner,turner2003escape} and towards predicting the optimal level of virulence in clinical infections of \textit{Mycobacterium tuberculosis} \cite{basu2009evolution}.

In this study, we apply these methods---formalized in an “evolutionary invasion analysis”---to examine two highly relevant viral pathogens: SARS-CoV-2 and hepatitis C virus (HCV), each informed by existing real-world empirical data that inform the parameter spaces. We chose these two systems because they represent contemporary epidemic scenarios defined by widely different disease ecologies and natural histories. SARS-CoV-2 dynamics are driven by direct transmission between those infected, via both symptomatic and asymptomatic transmission \cite{mizumoto2020estimating, nishiura2020estimation, kronbichler_asymptomatic_2020}. Hepatitis C virus, on the other hand, is largely transmitted indirectly between persons who inject drugs (PWID) in modern settings via drug equipment \cite{alter_hcv_2011}. 

Using evolutionary invasion analysis, we offer an integrated method for modeling the evolution of virulence across these two systems. In SARS-CoV-2, we learn that two different conceptual framings of virulence, one involving virulence as a function of the transmission from symptomatic individuals and another where it is a function of both symptomatic and asymptomatic transmission lead to different evolutionary patterns. Examining a mathematical model of HCV in PWID identifies an ESS virulence level that depends on treatment rate, progression into late-stage disease, and self-clearance rate. 


Summarizing, we learn that there should be no singular expectation for how virulence will evolve in a population. More broadly, we reflect on our findings for the individual outbreak scenarios (COVID-19 and HCV) and how they may inform larger conversations surrounding how we measure, understand, and prognosticate the evolution of virulence in epidemics, with implications for mathematical epidemiology and public health. 


\section*{Methods}\label{sec:method}

\subsection*{On the choices of disease models} This study focuses on viral pathogens, as understanding and predicting how virulence evolves in these systems have been especially dubious. Dozens of examples could be used to examine this, but we chose two---SARS-CoV-2 and hepatitis C virus (HCV)---both contemporary public health concerns. In addition, they represent diseases with different natural histories, allowing us to examine ESS virulence evolution in many different settings. We emphasize that our goal is not to offer any particular argument or intervention but to examine virulence evolution in varied disease systems. 


\subsection*{Notes on the approach} In this study, we examine the impact of pathogen evolution on the parameters of a disease, such as virulence and transmissibility, through a mathematical model with two strains. We utilize a system of \glspl{ode} of the compartmental model to determine the fitness function of the strains. The analysis identifies the fitness function as the $\gls{r0}$, which depends on the evaluation parameters. By establishing the fitness function, we can gain insight into the pathogen's \glspl{ess} and analyze its sensitivity to other parameters. We present an algorithm for these calculations and apply it to the evaluation of \gls{covid} and \gls{hcv} using \gls{ode} models.

\subsection*{Notes on terminology}
``Strain" is the vernacular sometimes used in applications of the evolutionary stable strategy in pathogen evolution. In our study, we use the term ``strain" to mean different phenotypic variants of a viral pathogen still belonging to the same type. We recognize the dubiousness associated with how viruses are grouped (e.g., clone, population, quasispecies), but are using language that is consistent with others in related fields.  

Similarly, ``virulence" is a famously complicated term, often used to describe different phenotypic impacts of pathogen infection. For the purposes of our study, one might use a standard definition, related to the harm caused to the host on behalf of a pathogen's infection. However, because our study utilizes mathematical models, we try to be explicit and consistent about its definitions. We translate virulence as the rate of death from infection (``infected death rate"), as this captures the ultimate sort of harm caused by a pathogen. 

\subsection*{Model of disease dynamics with the evolution of pathogen}
\begin{figure}[ht]
    \centering
    \includegraphics[width=\textwidth]{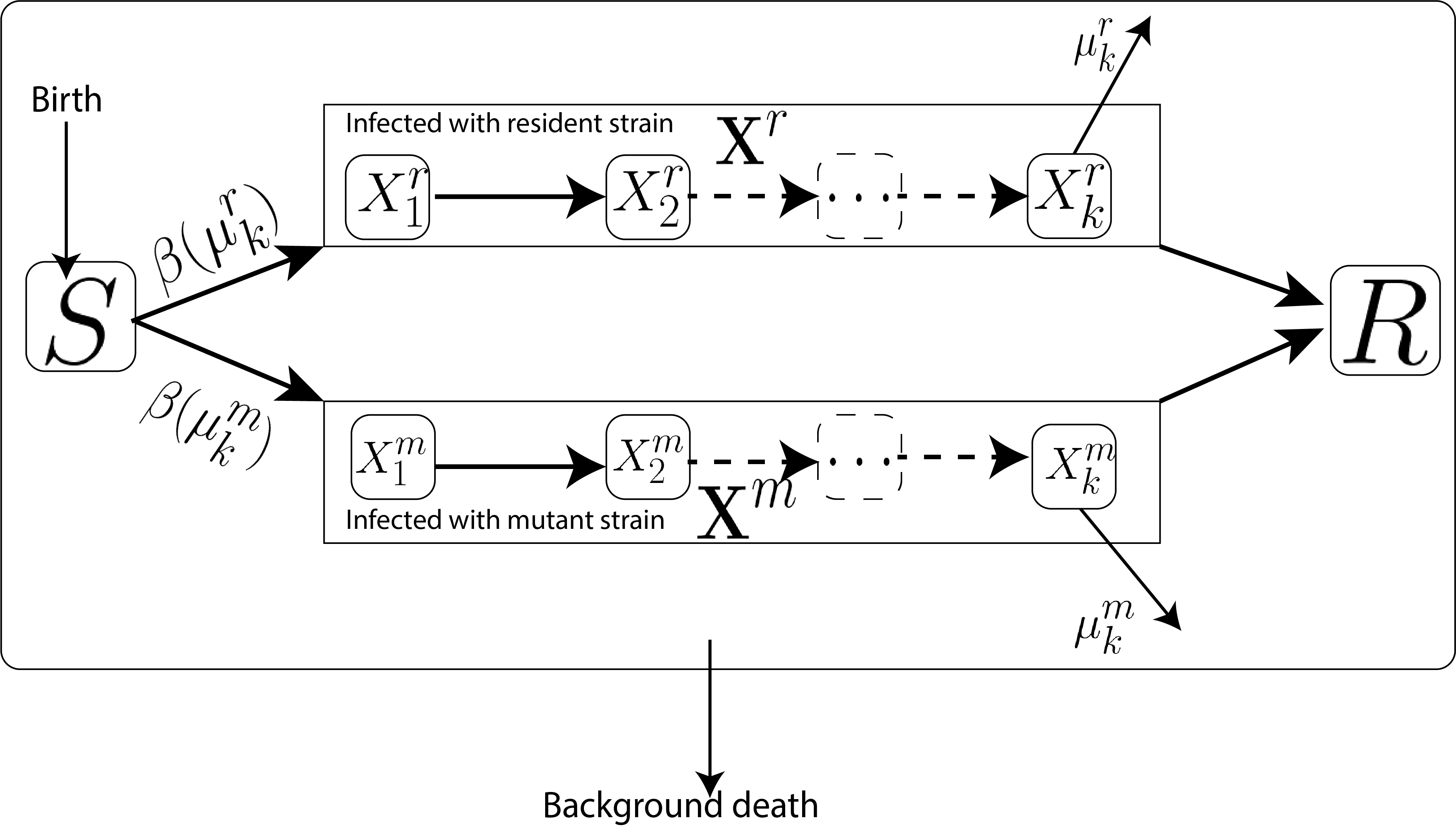}
    \caption{This figure shows the general compartment model with the assumption of no super-infections. It demonstrates the dynamics of disease with two pathogen strains. Pathogen evolution can impact virulence $\mu_k$ and transmissibility. The transmission rate, represented by $\beta$, is presumed to be a function of the virulence, $\mu_k$.}
    \label{fig:overview}
\end{figure}

The evolution of communicable disease pathogens can be simulated using a host compartmental \gls{ode} model. Here, we analyze a model that involves two viral strains and assumes that a host infected with the  resident pathogen (characterized by $r$) cannot simultaneously be infected with the invading mutant (characterized by $m$) and that a host immune to the resident pathogen is also immune to the mutant pathogen (no super-infections). Either strain can contaminate the susceptible compartment ($S$) and then proceeds through $k$ infected compartments ($\{X_{i_j}\}_{j=1}^k$ for $i\in\{r,m\}$) before reaching the recovery compartment ($R$) (as depicted in \cref{fig:overview}).

This study assumes that the pathogen's evolution influences the disease's virulence and transmission. Additionally, we assume that the transmission rate $\beta$ of fully infected hosts depends on virulence (as shown in \cref{fig:overview}) to derive a fitness function for the strains. The general \gls{ode} model is expressed as follows (\cref{eq:ODEsysOverview}).

\begin{equation}\label{eq:ODEsysOverview}
\begin{aligned}
    \frac{dS}{dt}&= b(N)-Sf_r\mathbf{X}_r-S f_m\mathbf{X}_m -d(S)\\
                            \frac{dR}{dt}&=g_r\mathbf{X}_r+ g_m \mathbf{X}_m-d(R)\\
        \frac{d\mathbf{X}_r}{dt}&= \mathbf{S}\mathbf{F}_r\mathbf{X}_r-\mathbf{D}_r\mathbf{X}_r \\
                \frac{d\mathbf{X}_m}{dt}&= \mathbf{S}\mathbf{F}_m\mathbf{X}_m-\mathbf{D}_m\mathbf{X}_m
\end{aligned}
\end{equation}
where the notation is explained in \cref{fig:overview} and \cref{tab:odeDis}. 

\begin{table}[ht]
\centering
\caption{This table provides a description of each variable and parameter in \cref{eq:ODEsysOverview}.}
\label{tab:odeDis}
\begin{tabular}{|p{4cm}|p{8cm}|}
\hline
\textbf{Notation} & \textbf{Description} \\ [0.5ex]
\hline\hline
 \gls{S}& \glsdesc{S}  \\
 \hline
 \gls{AdjS}& \glsdesc{AdjS}  \\
  \hline
 \gls{I}& \glsdesc{I}  \\
  \hline
  \gls{R}& \glsdesc{R}  \\
   \hline
 \gls{bN}& \glsdesc{bN}  \\
  \hline
  \gls{dA}& \glsdesc{dA}  \\
   \hline
\gls{fj}& \glsdesc{fj}  \\
 \hline
\gls{gj}& \glsdesc{gj}  \\
 \hline
\gls{Finj}& \glsdesc{Finj}  \\
 \hline
\gls{Dinj}& \glsdesc{Dinj}  \\
\hline
\end{tabular}
\end{table}

The principle that a disease-free equilibrium must be established as a requirement for disease models is widely recognized. This is because the transmission of the pathogen cannot occur without any initial cases of infection. Hence, the right-hand side of the infection groups in the general model, as outlined in \cref{eq:ODEsysOverview}, has the form $AX$, where $A$ is a matrix (whose entries may depend on the infection variables) and $X$ is the vector of infection groups. Furthermore, $A$ can be decomposed into the $\mathbf{S}\mathbf{F}-\mathbf{D}$ form. To achieve the basic reproduction number, $\gls{r0}$, biologically relevant values for $\gls{AdjS}\gls{Finj}$ and $\gls{Dinj}$ are chosen, satisfying the hypothesis of next-generation theory ($\gls{AdjS}\gls{Finj},\  \gls{Dinj}^{-1}>0$ and the spectral bound of $-\gls{Dinj}$ being less than zero) \cite{RN2}. The premise of ``no superinfections" implies that there is no direct interaction between competing strains, which can be further described as each strain's derivatives (\cref{eq:ODEsysOverview}) being determined exclusively by its own parameters and infection variables, along with the presence of susceptible variables. With these assumptions and explanations presented in \cref{tab:odeDis}, the system has the potential to exhibit four distinct equilibrium solutions, denoted as $\hat{V}_l=(\hat{S}_l,\hat{\mathbf{X}}_{r,l},\hat{\mathbf{X}}_{m,l},\hat{R}_l)$ for $l=1,\dots, 4$ (as outlined in \cref{table:fixedpoints}). The stability criteria for the equilibrium points can be expressed using a form of \gls{r0}. The \cref{remark:r0ForSingleVar} and \cref{remark:r0ForMultiVar} explain the \gls{r0} for the single-strain and two-strain \gls{ode} models, respectively. We utilize the notations $\mathbf{S^*},\ \mathbf{\bar{F}}_j,\ \mathbf{\bar{D}}_j$ for $j=r,m$ to denote the values of those matrices at the \gls{dfe}).

\begin{table}[ht]
\centering
\begin{tabular}{|p{3.5cm}|p{5.2cm}|p{4cm}|}
\hline
\textbf{Fixed Point} & \textbf{Description} & \textbf{Existence Condition} \\ [0.5ex]
\hline\hline
$\hat{V}_1=(S^*,0,0,0)$ & No infections & Always present in the model  \\
\hline
$\hat{V}_2=(\hat{S}_r,\hat{\mathbf{X}}_r,0,\hat{R}r)$ & Infected only with resident strain. We define this as \gls{mfe}. & $|\mathbf{S}\mathbf{F}_r-\mathbf{D}_r|_{\hat{V}_2}=0$ \\
\hline
$\hat{V}_3=(\hat{S}_m,0,\hat{\mathbf{X}}_m,\hat{R}m)$ & Infected only with mutant strain & $|\mathbf{S}\mathbf{F}_m-\mathbf{D}_m|_{\hat{V}_3}=0$ \\
\hline
$\hat{V}_4=(\hat{S},\mathbf{X}_r^*,\mathbf{X}_m^*,\hat{R})$ & Infected with resident or mutant strain (Co-infected) & $|\mathbf{S}\mathbf{F}_r-\mathbf{D}_r|_{\hat{V}4}=0$ and $|\mathbf{S}\mathbf{F}_m-\mathbf{D}_m|_{\hat{V}_4}=0$ \\
\hline
\end{tabular}
\caption{This table presents a list of equilibrium points in the mathematical model represented by the system of \glspl{ode} shown in \cref{eq:ODEsysOverview}. Understanding the properties of fixed points is essential for understanding the system's long-term behavior. In this computation, we leverage the fundamental mathematical theorem that states that if $x$ is a non-zero vector and $Ax=0$, then the matrix $A$ is singular, which means that its determinant is zero ($|A|=0$).}
\label{table:fixedpoints}
\end{table}

\begin{remark}[\glsfmttext{r0} for Single-Strain Model]\label{remark:r0ForSingleVar}
Suppose a single-strain model is explained by the reduced system of \glspl{ode} from equation \eqref{eq:ODEsysOverview}. Then, the \gls{r0} is given by: 
\begin{equation}
    \mathcal{R}_0 = \rho(\mathbf{S^*\bar{F}}_r\mathbf{\bar{D}}_r^{-1})
\end{equation}
where $\rho(.)$ represents the spectral radius of a given matrix. Note that, the eigenvalues of a block upper triangular matrix of the form determine the stability condition of the \gls{dfe} 
 \begin{align}
     J(S^*)=\begin{pmatrix}
     J_* & M_r \\
     0 & \mathbf{S^*\bar{F}}_r-\mathbf{\bar{D}_r}
     \end{pmatrix}
 \end{align}
 where $J_*,\ M_r$ are some matrices corresponding to the \glspl{ode} of susceptibles and recovered compartments. Therefore, the invasion of the pathogen depends on the sign of the maximum eigenvalue of $\mathbf{S^*\bar{F}}_r-\mathbf{\bar{D}_r}$. This can be expressed in terms of the condition of \gls{r0} (a detailed explanation can be found in \cite{RN1}  and \cite{RN2}).
\end{remark}

\begin{remark}[\glsfmttext{r0} for Two-Strain Model]\label{remark:r0ForMultiVar}
Suppose a two-strain model is explained by the equation \eqref{eq:ODEsysOverview}. Then, the $\gls{r0}$ for two-strain model is given by: 
\begin{equation}
    \gls{r0} = \max_{i=r,m}{\rho(\mathbf{S^*\bar{F}}_i\mathbf{\bar{D}}_i^{-1})}.
\end{equation}
In this case, the stability of the \gls{dfe} can be determined by the following Jacobian matrix (block upper triangular form):
 \begin{align}
     J(S^*)=\begin{pmatrix}
     J_* & M_r & M_m \\
     0 & \mathbf{S^*}\mathbf{\bar{F}}_r-\mathbf{\bar{D}_r} & 0\\
     0 & 0& \mathbf{S^*}\mathbf{\bar{F}}_m-\mathbf{\bar{D}_m} 
     \end{pmatrix}.
 \end{align}
 where $J_*,\ M_r, \ M_m$ are some matrices corresponding to the \glspl{ode} of susceptibles and recovered compartments. Therefore, \gls{dfe} is locally asymptotically stable if $\gls{r0}<1$, but unstable if $\gls{r0}>1$.
\end{remark}

\subsection*{Fitness function}
To calculate the fitness function of the mutant, a stability analysis of the endemic equilibrium $\hat{V}_2$ of the resident population can be performed. This approach considers the mutant as the solely infected compartment, with the resident population being treated as uninfected. The fitness function can be effectively analyzed by utilizing the next-generation matrix theory at the mutant-free equilibrium. Notice that, Jacobin matrix of the system (\cref{eq:ODEsysOverview}) at $\hat{V}_2$ is given by the form:
 \begin{align}
     J(\hat{V}_2)=\begin{pmatrix}
     J_2 & M_2 \\
     0& J_m=\mathbf{\hat{S}}_r\mathbf{\bar{F}}_m-\mathbf{\bar{D}}_m 
     \end{pmatrix}
 \end{align}
 where $\mathbf{\hat{S}}_r$ is the value of $\mathbf{S}$ at \gls{mfe} and  $J_2, M_2$ are some matrices corresponding to the \glspl{ode} for compartments of susceptibles, recovered and infected with the resident strain.
Accordingly, it is established that the matrix $J(\hat{V}_2)$ has a block upper triangular form, consisting of sub-matrices $J_2$ and $J_m$. The eigenvalues of $J(\hat{V}_2)$ are the same as those of $J_2$ and $J_m$.

Based on this observation, a crucial criterion for the successful invasion of the mutant strain can be deduced. Specifically, the matrix $J_m$ must possess at least one eigenvalue with a positive real part. This condition, called the \gls{r0} requirement for the mutant invasion, is formally stated as \cref{lemma:mut}.

\begin{lemma}[\gls{r0} for pathogen mutant invasion]\label{lemma:mut}Let the model for an infectious disease be represented by the system of \glspl{ode} in \cref{eq:ODEsysOverview}. Suppose \gls{mfe} is a locally asymptotically stable equilibrium solution for \cref{eq:ODEsysOverview} restricted to resident strain. Consider the basic reproduction number for mutant invasion:
\begin{equation}\label{eq:RoMFE}
    \gls{r0}(\hat{V}_2)=\rho(\mathbf{\hat{S}}_r\mathbf{\bar{F}}_m\mathbf{\bar{D}}_m^{-1}).
\end{equation}
 Then the \gls{mfe} is locally asymptotically stable if $\gls{r0}(\hat{V}_2)<1$, and unstable if $\gls{r0}(\hat{V}_2)>1$.
\end{lemma}

Consequently, the calculation of $\gls{r0}(\hat{V}_2)$ constitutes the initial step in the process of determining a fitness function for the resident strain. It is important to note that the $\gls{r0}(\hat{V}_2)$ can be simplified as follows:
\begin{equation}\label{eq:r0MFEval}
\begin{aligned}
        \gls{r0}(\hat{V}_2) &=\frac{\phi_m}{\phi_r} 
\end{aligned}
\end{equation}
 where   $\phi_j =\gls{r0}_j\prod_{i=2}^n \Big|\frac{\lambda_i(\mathbf{S}^*\bar{\mathbf{F}}_j\bar{\mathbf{D}}_j)}{\lambda_i(\mathbf{\hat{S}}_r\bar{\mathbf{F}}_j\bar{\mathbf{D}}_j)}\Big|$  for  $j=r,m$ and $\lambda_i(.)$ denote the $i^{th}$ eigenvalue of a given matrix. This result can be achieved by utilizing the fact that the determinant of a matrix is equal to the product of its eigenvalues. Thus, the fitness function for the strains is defined as follows:
\begin{align}
\Phi(\mu_k^r,\mu_k^m)= \mathcal{R}_0(\hat{V}_2) -1,
\end{align}
where the mutant will invade if and only if $\Phi(\mu_k^r,\mu_k^m)>0$. It is worth mentioning that different choices for the fitness function exist, such as $\Phi_1(\mu_k^r,\mu_k^m)=\phi_m-\phi_r$. However, the \gls{r0} value for the disease models can be easily obtained in epidemiology, making it a suitable choice for our calculations. Examples will be provided in subsequent sections.

We aim to identify the virulence level that maximizes the invasion's fitness. In other words, we are interested in finding a strain that a mutant cannot invade. For discussion, we will assume $\mathcal{R}_0(\hat{V}_2)=\frac{\phi(\mu_k^m)}{\phi(\mu_k^r)}$, which is consistent with practical applications and \cref{eq:r0MFEval}. Under this assumption, the optimal fitness is achieved when $\mu_k^*=\argmax{\phi(\mu_k)}$. This value represents the \glsentrydesc{ess} (\gls{ess}) for the pathogen.

\begin{figure}[ht]
    \centering
         \begin{subfigure}[b]{0.48\textwidth}
             \centering
    \includegraphics[width=\textwidth]{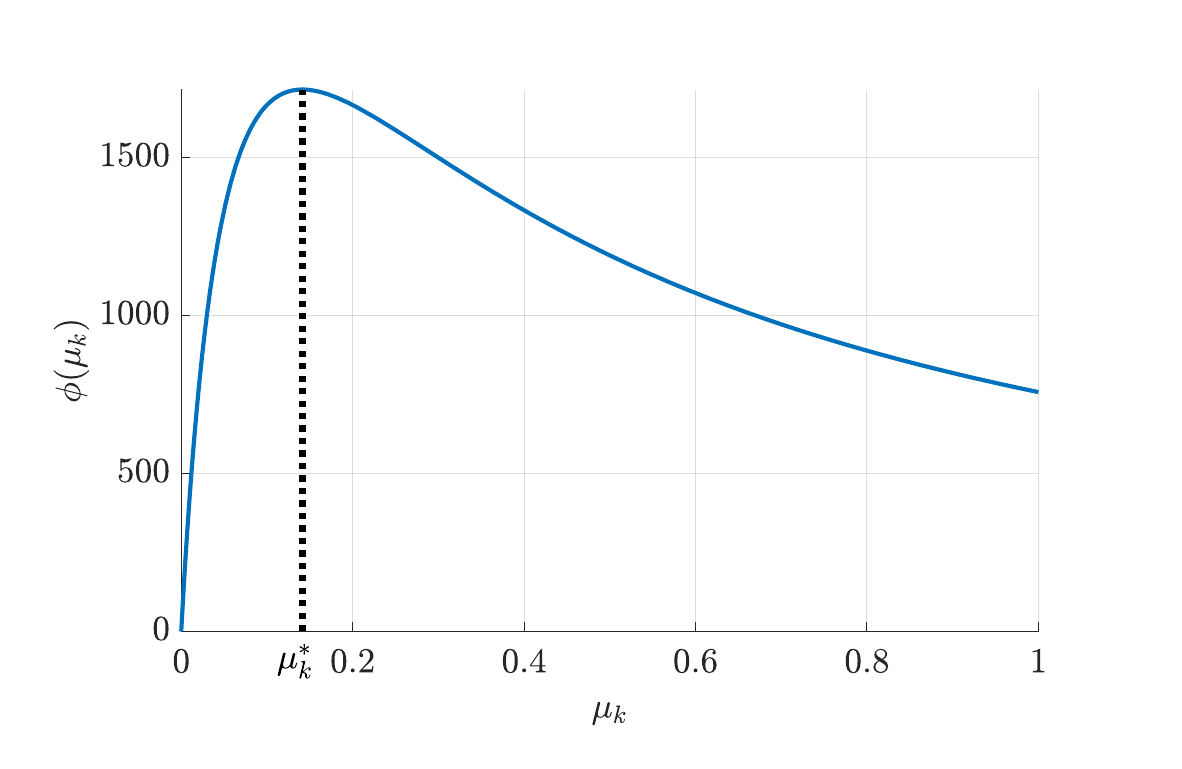}
    \caption{}
    \label{fig:phiFun}
         \end{subfigure}
     \hfill
    \begin{subfigure}[b]{0.48\textwidth}
         \centering
\includegraphics[width=\textwidth]{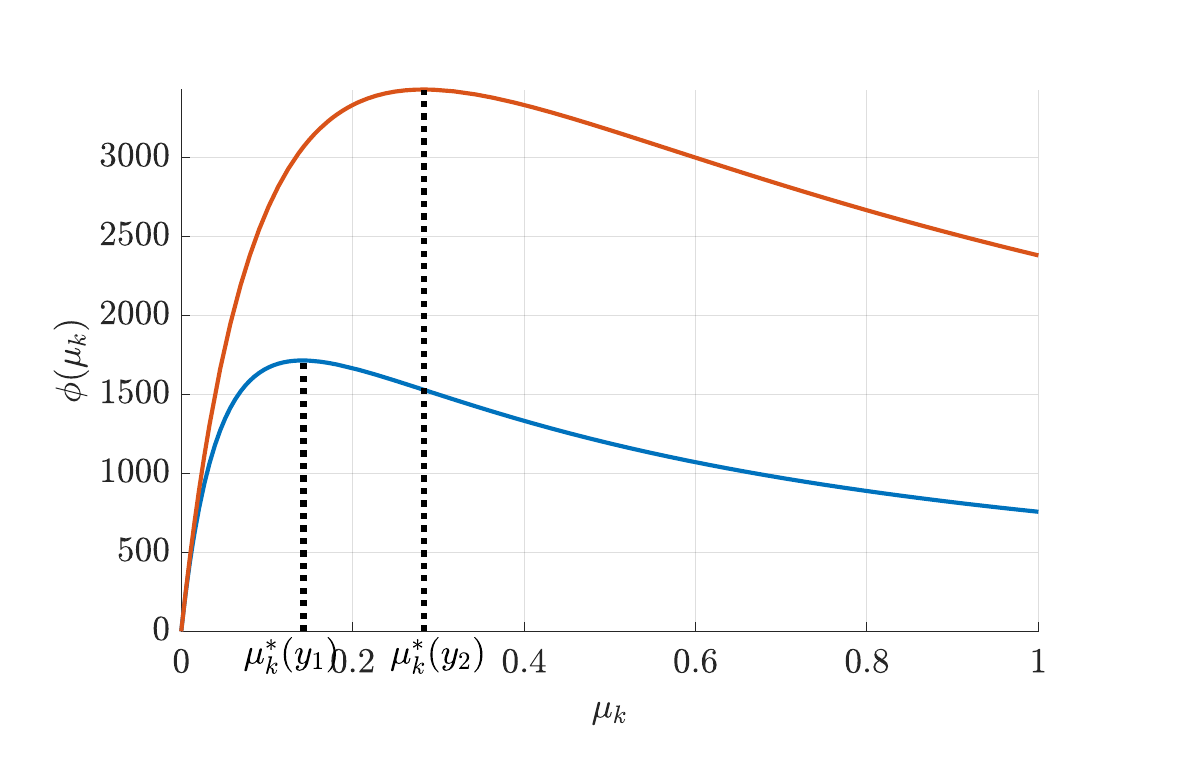}
    \caption{}
    \label{fig:EssWithPara}
     \end{subfigure}
         \caption{\textit{Fitness function and evolutionary stable strategy for a virus population.} Figure (a) shows the fitness function with respect to  virulence for a given strain, assuming no direct dependence from others. Figure (b) illustrates the \gls{ess} change with a given parameter $y$. }
\end{figure}

\subsection*{Evolutionary stable strategies}
Determining an \glspl{ess} for a pathogen is crucial in understanding the dynamics of pathogen populations. The concept of \gls{ess} is based on the premise that if a current pathogen population with specific parameters is not vulnerable to invasion by a mutant, then these parameters represent a stable strategy for the pathogen. The fitness function, $\phi(\mu_k)$, is used to analyze these strategies. As previously discussed, the \gls{ess} is given by $\mu_k^*=\argmax{\phi(\mu_k)}$. This can be demonstrated through \cref{fig:phiFun} where it can be seen that a pathogen with a virulence value of $\mu_k^r=\mu_k^*$; cannot be invaded by a mutant as its fitness is at its maximum. If the fitness function, $\phi(\mu_k)$, is continuous on the closed interval $[0,1]$, then we can determine the global \gls{ess} through the calculation of the absolute maximum value.

Moreover, if $\phi(\mu_k)\in C[0,1]$ is twice differentiable at $\mu_k^*$, derivative tests can be utilized to find the local/absolute maximum values and \gls{ess}. It can be concluded that if $\phi(\mu_k)$ is twice differentiable at $\mu_k^*$, then $\mu_k^*$ is an \gls{ess} if and only if
\begin{align}
    \frac{d\phi}{d\mu_k}\Big|_{\mu_k=\mu_k^*}&=0,~~~ \text{ and } ~~~
     \frac{d^2\phi}{d\mu_k^2}\Big|_{\mu_k=\mu_k^*}< 0 .
\end{align}

\subsection*{Sensitivity analysis of \glsentrytext{ess}}
After determining the \gls{ess}, it is beneficial to analyze the sensitivity of the \gls{ess} with regard to other parameters (as illustrated in \cref{fig:EssWithPara}). For any given parameter $y$, the change in the \gls{ess} $\mu_k^*$ with respect to the increase of $y$ can be determined by calculating $\frac{d\mu_k^*}{dy}$. In order to do so, let $\psi(x,y)=\frac{\partial \phi(\mu_k,y)}{\partial \mu_k}\Big|_{\mu_k=x}$ and let $\mu_k^*$ be a function of $y$. It is important to note that $\mu_k^*$ satisfies the following equation:
\begin{align}\label{eq:sens1}
    \frac{\partial}{\partial y}\psi(\mu_k^*,y)= \frac{\partial \psi(\mu_k,y)}{\partial \mu_k}\Big|_{\mu_k=\mu_k^*}\frac{d \mu_k^*}{d y}+ \frac{\partial \psi(\mu_k,y)}{\partial y}\Big|_{\mu_k=\mu_k^*}=0
\end{align}
and it can be rewritten as:
\begin{equation}\label{eq:inva}
    \frac{d \mu_k^*}{d y} = - \frac{\frac{\partial^2 \phi(\mu_k,y)}{\partial y \partial \mu_k}\Big|_{\mu_k=\mu_k^*}}{\frac{\partial^2 \phi(\mu_k,y)}{\partial^2 \mu_k}\Big|_{\mu_k=\mu_k^*}}\propto \frac{\partial^2 \phi(\mu_k,y)}{\partial y \partial \mu_k}\Big|_{\mu_k=\mu_k^*}.
\end{equation}
Consequently, by examining the sign of the partial derivative of $\phi(\mu_k,y)$ with respect to the parameter $y$ at $\mu_k=\mu_k^*$, one can determine the sensitivity of the \gls{ess} with respect to the change in $y$. This allows for an analysis of the behavior of the \gls{ess} with respect to any given parameter without the need for an explicit calculation of $\mu_k^*$. This calculation can further be simplified with the given problem. For example, let $\phi(\mu_k, y)=\frac{U(\mu_k,y)}{W(\mu_k, y)}$ where $U, W>0$. This results in the following expression:
\begin{align}
\frac{\partial^2 \phi(\mu_k,y)}{\partial y \partial \mu_k}\Big|_{\mu_k=\mu_k^*} \propto \frac{\partial}{\partial y}\big( W \frac{\partial U}{\partial \mu_k}- U \frac{\partial W}{\partial \mu_k} \big)\Big|_{\mu_k=\mu_k^*}
\end{align}
and, the sign of $\frac{d \mu_k^*}{d y}$ is equal to the sign of $\frac{\partial}{\partial y}\big( W \frac{\partial U}{\partial \mu_k}- U \frac{\partial W}{\partial \mu_k} \big)\Big|_{\mu_k=\mu_k^*}$.

The calculation of the \gls{ess} and the analysis of changes in the ESS with respect to parameters can be summarized in a structured framework. This framework provides a systematic approach for analyzing the stability of ESS and its sensitivity to different parameters, which is crucial in understanding the behavior of the pathogen population.

\begin{myframe}{\textbf{Box 1.} Framework for Calculating \glsentrytext{ess} and Conducting Sensitivity Analysis}\label{fram:ESS}
    \begin{myenumerate}
        \item Use the \gls{ode} model (SIR, SEIR, etc.) and calculate the basic reproduction number \gls{r0}. (This is just the currently used mathematical model for the diseases assuming only one strain for the pathogen.)
        \item Choose a base model parameter (decision variable) defined by evolution. (In this study, we consider the mortality rate due to the disease (virulence) as that parameter.)
        \item You may categorize the model parameters into categories such that the parameters that are 
        \begin{myitemize}
        \item independent of the evolution,
    \item dependent on the evolution but not directly related to the decision variable, or
    \item function of the decision variable. (For example, consider the transmission rate $\beta$ as a function of virulence $\mu_k$).
    \end{myitemize}
   
    \item Find \gls{mfe}, $\hat{V}_2$ and evaluate the $\gls{r0}(\hat{V}_2)=\frac{\phi_m}{\phi_r} $. 

\item Considers $\phi(x)=\phi_{x}$ (where $x$ depends on the given strain) as a function of the chosen decision variable and finds maxima that provide the ESS.
\item Finally, consider the equation $\phi(x,y) = \phi_x(y)$ to determine the sensitivity of ESS to the parameter $y$.
     \end{myenumerate}
\end{myframe}

The next section will demonstrate the proposed framework through several examples.

\section*{Results from illustrative cases}
In this section, we provide a demonstration of the theoretical framework for evaluating Evolutionary Stable Strategy (ESS), as described in the previous section, 
using COVID-19 and HCV as case studies. 

\subsection*{SEIR  model type: \gls{covid} Example}
\begin{figure}[ht]
    \centering
    \includegraphics[width=\textwidth]{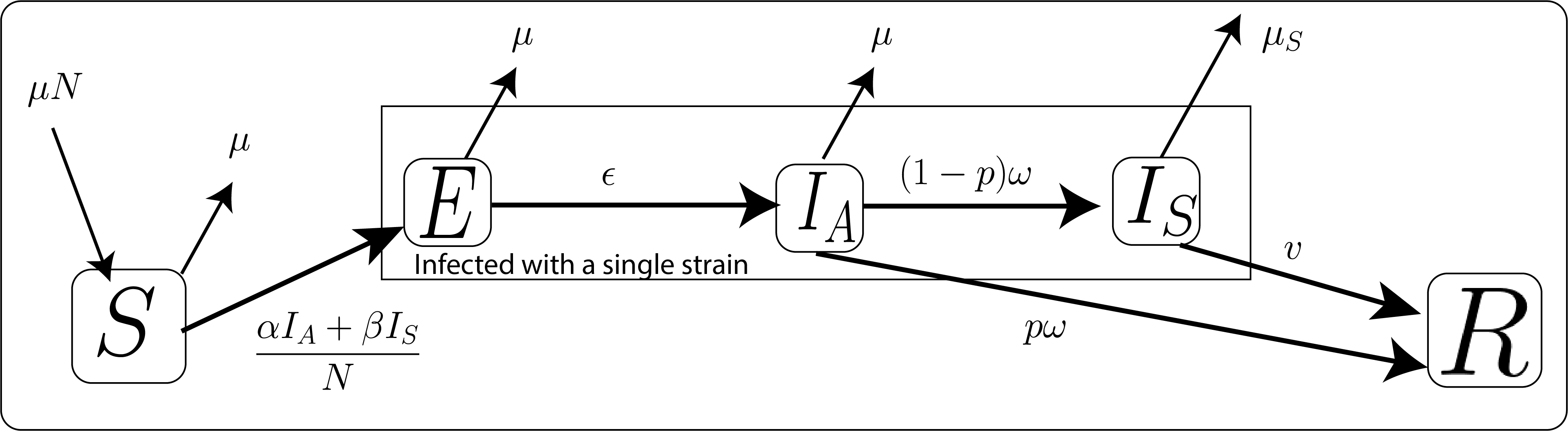}
    \caption{This compartmental model diagram illustrates the system dynamics of the susceptible, exposed, asymptomatic, symptomatic, and recovered compartments in relation to \gls{covid}. Further explanations of the variables and parameters utilized in this figure can be found in \cref{tab:CovidVara} and  \cref{tab:CovidPara}. The system of \glspl{ode} governing this model is presented in \cref{eq:covidModel}.}
    \label{fig:CovidModel}
\end{figure}

We explore the Susceptible-Exposed-Infected-Removed (SEIR) compartment model to showcase the proposed theory, with the spread of strains of the \gls{covid} as an example. We use the equations presented in \cref{eq:covidModel} to model the spread of a specific virus strain in the host population (refer to \cite{ogbunugafor2020variation} for details on the single-strain model). \cref{fig:CovidModel} visualizes the SEIR model.

\begin{equation}\label{eq:covidModel}
    \begin{aligned}
        \frac{d S}{dt} &= \mu (N-S) -\big( \frac{\alpha I_A + \beta I_S}{N}\big)S\\
        \frac{d E}{dt} &= \big( \frac{\alpha I_A + \beta I_S}{N}\big)S-(\epsilon + \mu)E \\
        \frac{d I_A}{dt} &= \epsilon E -(\omega + \mu) I_A\\
        \frac{d I_S}{dt} &= (1-p)\omega I_A -(v + \mu_s) I_S\\
        \frac{d R}{dt} &= p\omega I_A+v I_S - \mu R\\
    \end{aligned}
\end{equation}

\begin{table}[ht]
\centering
\caption{This table describes variables in \cref{eq:covidModel}. All variables are measured as the number of people.}
\label{tab:CovidVara}
\begin{tabular}{p{2cm}p{4cm}}
\toprule
\rowcolor{gray!80}
\textbf{Variables} & \textbf{Description}\\
\midrule
\rowcolor{white}
$N$ & Total population \\
\rowcolor{gray!30}
$S$ & Susceptible individuals \\
\rowcolor{white}
$E$ & Exposed individuals \\
\rowcolor{gray!30}
$I_A$ & Asymptomatic individuals \\
\rowcolor{white}
$I_S$ & Symptomatic individuals  \\
\rowcolor{gray!30}
$R$ & Recovered individuals \\
\bottomrule
\end{tabular}
\end{table}

\begin{table}[ht]
\centering
\caption{This table describes parameters in  \cref{eq:covidModel}. The infected death rate is considered as the measure of virulence in the model, and it is highlighted in the table as a row of bold text.}
\label{tab:CovidPara}
\begin{tabular}{p{2cm}p{7.3cm}p{1.2cm}}
\toprule
\rowcolor{gray!80}
\textbf{Parameters} & \textbf{Description} & \textbf{Units} \\
\midrule
\rowcolor{white}
$\mu$ & Natural death rate  & $\si{day}^{-1}$  \\
\rowcolor{gray!30}
$\boldsymbol{\mu_S}$ & \textbf{Virulence (Infected death rate)} &$\boldsymbol{\si{\textbf{day}}^{-1}}$  \\
\rowcolor{white}
$\omega^{-1}$ & Expected time in the asymptomatic state  & $\si{days}$  \\
\rowcolor{gray!30}
$v$ & Recovery rate  & $\si{day}^{-1}$  \\
\rowcolor{white}
$p$ & The fraction that moves along the “mild” recovery track &  \\
\rowcolor{gray!30}
$\epsilon^{-1}$ & Average number of days before infectious & $\si{days}$  \\
\rowcolor{white}
$\alpha$ & Transmission rate through the asymptomatic individuals  & $\si{day}^{-1}$  \\
\rowcolor{gray!30}
$\beta$ & Transmission rate through the symptomatic individuals  & $\si{day}^{-1}$  \\

\bottomrule
\end{tabular}
\end{table}

Now, consider a two-strain scenario for the \gls{covid} model, where the strains are denoted as $r$ (resident) and $m$ (mutant). We extend the model by introducing the decision variable $\mu_S$ and assuming that the transmission rate $\beta$ is a function of $\mu_S$, denoted by $\beta_j=\beta(\mu_S^j)$ for $j=r,\ m$. Additionally, we assume that other parameters remain unchanged by evolution. To conform with the notation introduced in the theoretical framework, we use the notation $\mathbf{X}j=(E_j\ I{A_j}\ I_{S_j})^T$. In this context, the matrices $\mathbf{S}\mathbf{F}_j$ and $D_j$ relevant to this problem are given by:
\begin{equation*}
    \begin{aligned}
        \mathbf{S}\mathbf{F}_j=\begin{pmatrix}
            0 &\frac{\alpha S}{N} &\frac{\beta_j S}{N}\\
            0& 0 & 0 \\
            0 &0& 0
        \end{pmatrix}, \  
        \mathbf{D}_j=\begin{pmatrix}
            \epsilon+\mu &0 &0\\
            -\epsilon  & \omega + \mu & 0 \\
            0 &-(1-p)\omega  & v+\mu_S^j
        \end{pmatrix} \ \text{for} \ j=r,m. 
    \end{aligned}
\end{equation*}

Now, consider the case of resident strain at the mutant-free equilibrium $\hat{V}_r$. It can be observed that $|\mathbf{S}\mathbf{F}_r-\mathbf{D}r|_{\hat{V}_r}=0$, indicating that the following expression holds true:
\begin{equation}\label{eq:CovidSr}
\hat{S}_r= \frac{N(\epsilon+\mu)(\omega + \mu)(v+\mu_S^r)}{\epsilon\big(\alpha(v+\mu_s^r)+\beta_r\omega(1-p)\big)}.
\end{equation}
Furthermore, it should be noted that the basic reproduction number for the single-strain ($r$) model is given by:
\begin{equation}\label{eq:covidR0}
\rho(\mathbf{S}^*\mathbf{\bar{F}}_r\mathbf{\bar{D}}_r^{-1})=\gls{r0}_r=\frac{S^*\epsilon\big(\alpha(v+\mu_s^r)+\beta_r\omega(1-p)\big)}{N(\epsilon+\mu)(\omega+\mu)(v+\mu_s^r)}.
\end{equation}
where $S^*$ is the susceptible population at \gls{dfe}. Therefore, the basic reproduction number for the \gls{mfe} of the resident strain can be derived using the following equation:
\begin{equation} \label{eq:covidR0DFE}
\rho(\mathbf{\hat{S}}_r\mathbf{\bar{F}}_m\mathbf{\bar{D}}_m^{-1})=\gls{r0}(\hat{V}_2)=\frac{\hat{S}_r\epsilon\big(\alpha(v+\mu_s^m)+\beta_r\omega(1-p)\big)}{N(\epsilon+\mu)(\omega+\mu)(v+\mu_s^m)}.
\end{equation}
By substituting \cref{eq:CovidSr} into the expression for the basic reproduction number at \gls{mfe}, we obtain a form:
\begin{equation} \label{eq:covidR0MFE}
\gls{r0}(\hat{V}_2)=\frac{\Phi_m}{\Phi_r}
\end{equation}
where $\Phi_j=\gls{r0}_j$. This means that the available basic reproduction number information, modified with the parameters specific to evolution, can be directly used to determine the \gls{ess} for models in the SEIR setup. In the event where only $\mu_S$ and $\beta$ parameters are subject to evolution, the expression for $\gls{r0}(\hat{V}_2)$ can be simplified to the form $\gls{r0}(\hat{V}_2)=\frac{\phi(\mu_S^m)}{\phi(\mu_S^r)}$ where the function $\phi(\mu_S)=C_0 + C_1 \frac{\beta(\mu_s)}{v+\mu_s}$, and $C_0=\alpha$ and $C_1=\omega(1-p)$ are constants. 

Moreover, the maximum value of $\phi$ with respect to $\mu_s$ may attain at $\mu_S^*$ when the derivative of $\beta$ with respect to $\mu_S$ is equal to $\frac{\beta(\mu_S^*)}{v+\mu_S^*}$:
\begin{equation}
   \frac{d \beta}{d \mu_S} \big|_{\mu_S=\mu_S^*}=\frac{\beta(\mu_S^*)}{v+\mu_S^*}.
\end{equation}
It is important to note that if we use the basic reproduction number $\Phi=\gls{r0}$ instead of $\phi$ for the analysis, we will obtain the same results since $\Phi=C_2\phi$ with constant $C_2=\frac{S^*\epsilon}{N(\epsilon+\mu)(\omega + \mu)}$. However, to explicitly find the \glspl{ess}, we need to model the exact function $\beta(\mu_S)$. In this article, we demonstrate the concept using example functions such as $\beta(\mu_S)=\frac{\mu_S}{a_1+\mu_S},\ \tanh^2(a_1\mu_S+a_2),\ \sech^2(a_1\mu_S+a_2)$ and $\sin(a_1\mu_S+a_2)$, where $a_1$ and $a_2$ are constants. (Note that if $\beta$ is a decreasing function, the optimal solution is trivial $\mu_S=0$, so we exclude decreasing functions in this demonstration.) While we do not discuss estimating the $\beta$ function using data in this article, one can easily use any curve-fitting algorithm to identify the transmission function $\beta(\mu_S)$. For example, the theoretical analysis conducted by \citeauthor{massad_transmission_1987} \cite{massad_transmission_1987} utilized data from myxoma viruses to fit the hyperbolic secant squared function to demonstrate the relationship between transmission rate and virulence.  

\begin{figure}[ht]
    \centering
         \begin{subfigure}[b]{0.45\textwidth}
         \centering
         \includegraphics[width=\textwidth]{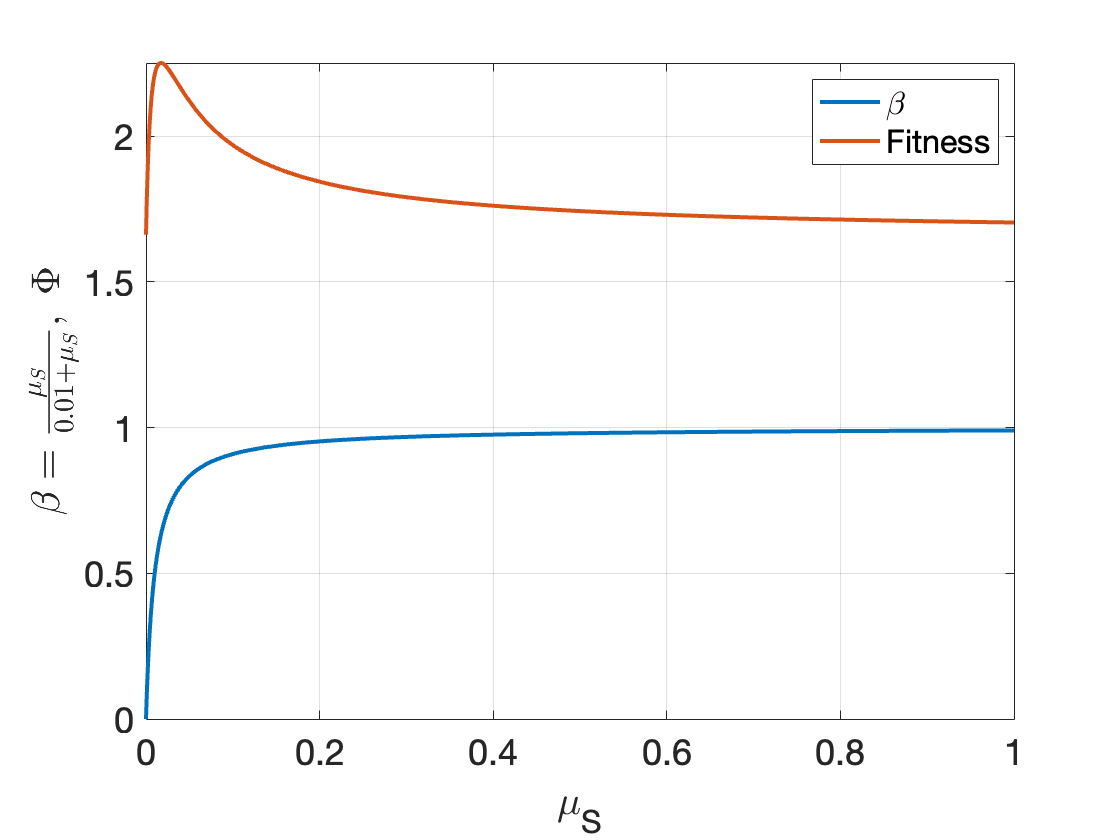}
         \caption{}
     \end{subfigure}
     \hfill
    \begin{subfigure}[b]{0.45\textwidth}
         \centering
         \includegraphics[width=\textwidth]{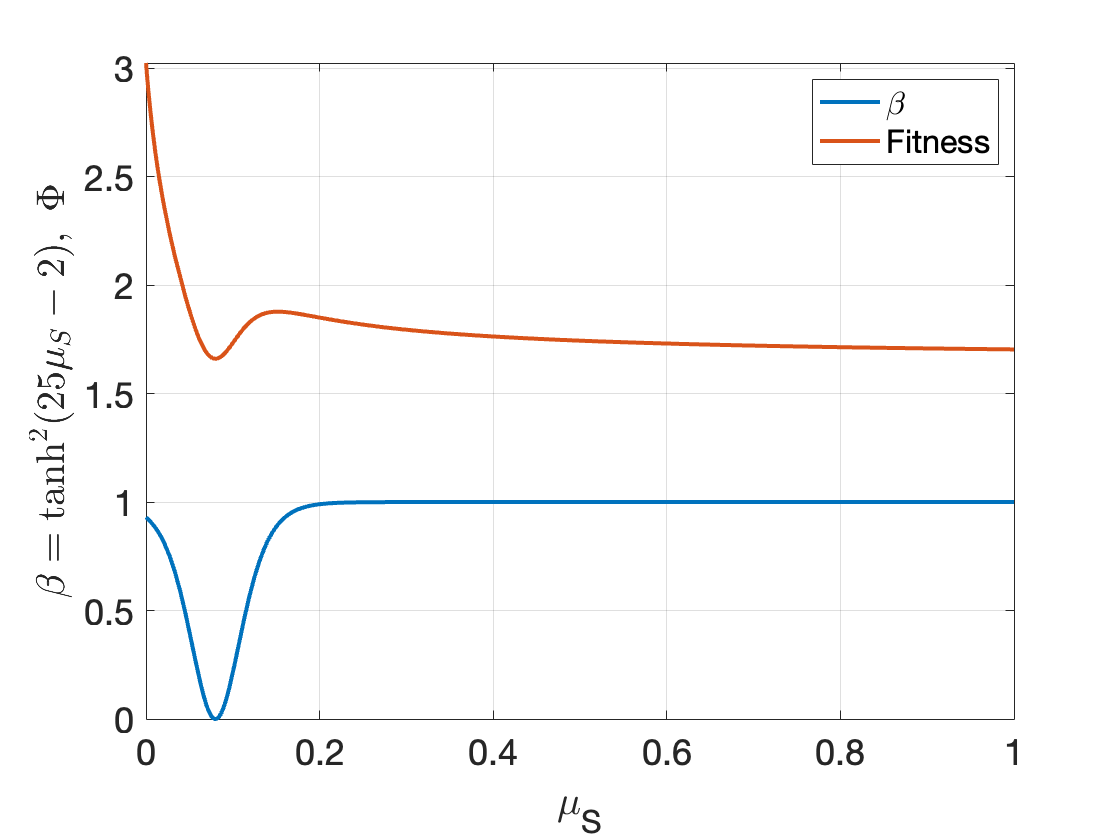}
         \caption{}
     \end{subfigure}
     \hfill
    \\
     \begin{subfigure}[b]{0.45\textwidth}
         \centering
         \includegraphics[width=\textwidth]{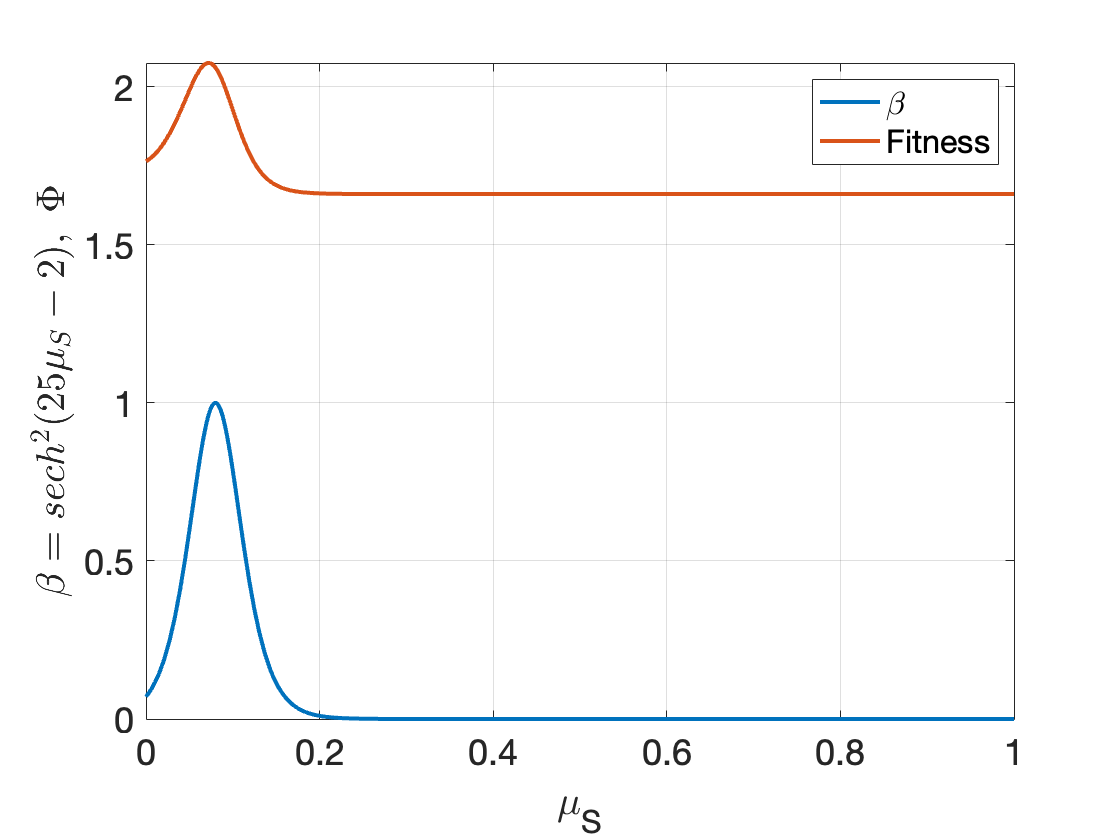}
         \caption{}
     \end{subfigure}
     \hfill
    \begin{subfigure}[b]{0.45\textwidth}
         \centering
         \includegraphics[width=\textwidth]{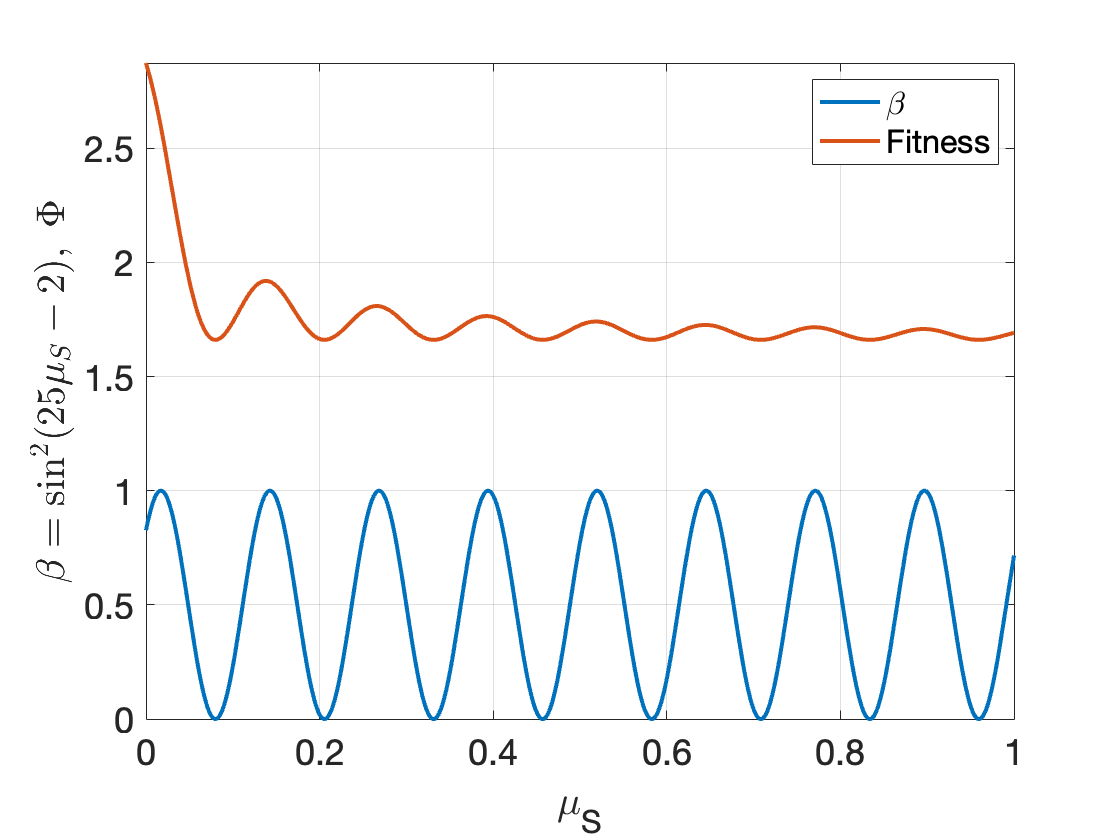}
         \caption{}
     \end{subfigure}
    \caption{The figure demonstrates the behavior of the fitness function $\Phi=\mathcal{R}_0$ when transmission function $\beta(\mu_S)$ equals to (a) $\frac{\mu_S}{0.01+\mu_S}$, (b) $\tanh^2(25\mu_S-2)$, (c)$\sech^2(25\mu_S-2)$ and (d) $\sin^2(25\mu_S-2)$. It can be observed that only (a) and (c) have a non-zero global \gls{ess}, while (b) and (d) have non-zero local \glspl{ess}.}
    \label{fig:EssWithMuBeta}
\end{figure}

Depending on the transmission function, there may be one or more local \glspl{ess} (see \cref{fig:EssWithMuBeta}). If $\beta$ is an oscillatory function, the fitness function may also be oscillatory and have multiple local ESS. Since $\mu_S$ is in the closed interval $[0,1]$ (and assuming that $\beta$ is bounded on $[0,1]$), the absolute maximum of the fitness function can be attained at either the endpoints of the interval $[0,1]$ or at a local maximum of the fitness function. \cref{fig:EssWithMuBeta} demonstrates the local and absolute maximums of the fitness function, which gives ESS for the pathogen. When a pathogen's fitness is at the local maximum of its fitness function, it may need to have a relatively significant change in virulence to attain global ESS. We leave it to the reader to choose the transmission function that best suits their project, and they can expose, extend, and justify their choice using our theoretical basis. For further discussion, we choose $\beta(\mu_S)=\sech^2(a_1\mu_S+a_2)$, but the same analysis can be conducted with any other function.

\begin{figure}[ht]
    \centering
         \begin{subfigure}[b]{0.45\textwidth}
         \centering
         \includegraphics[width=\textwidth]{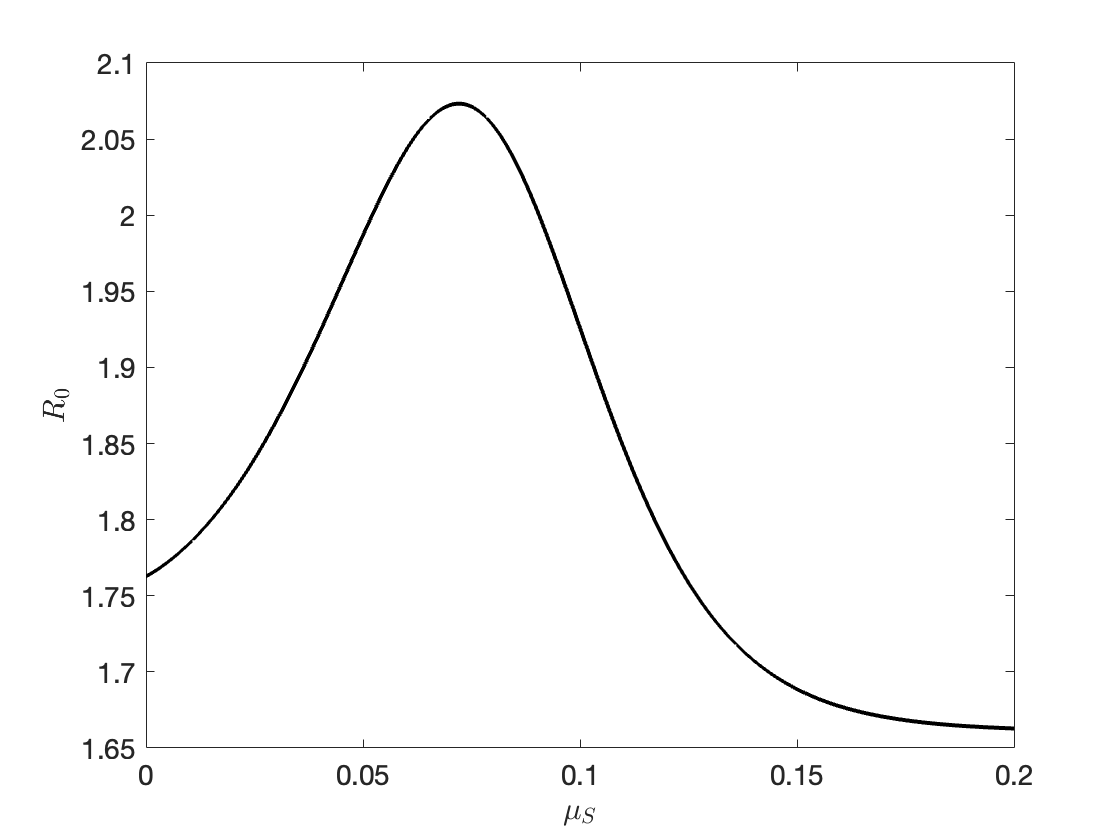}
         \caption{}
     \end{subfigure}
     \hfill
    \begin{subfigure}[b]{0.45\textwidth}
         \centering
         \includegraphics[width=\textwidth]{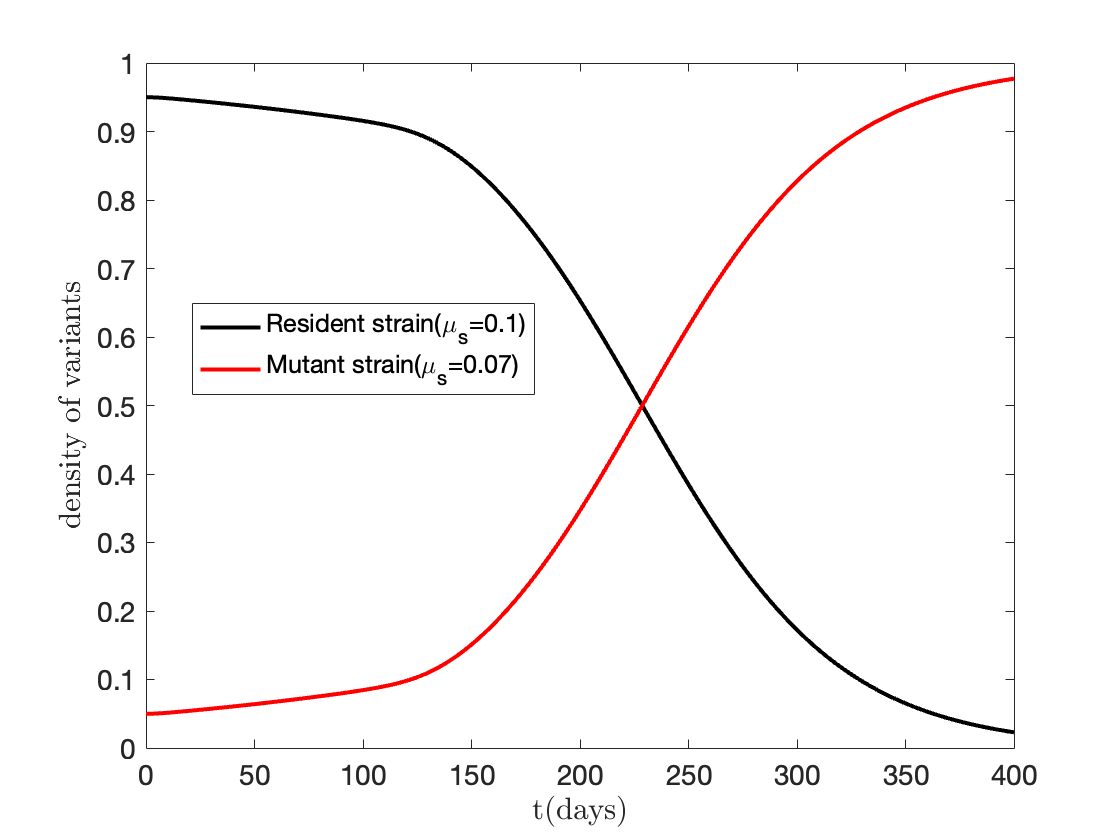}
         \caption{}
     \end{subfigure}

    \caption{This demonstrates the behavior of the \gls{r0} and the host density of each strain ($r,\ m$) for the \gls{covid} model with transmission function $\beta=\sech^2(25\mu_S-2)$ and parameter values: $\mu=0.000034, \omega^{-1}=3.119, v=0.031, p=0.956, \epsilon^{-1}=2.381,$ and $\alpha=0.429$ (we use these fixed parameter values, referring to \cite{ogbunugafor2020variation}). Panel (a) shows the variation of \gls{r0} as a function of virulence, and panel (b) shows host population density for both resident ($\mu_S=0.1$) and mutant strains ($\mu_S=0.07$). It is worth noting that the maximum value of \gls{r0} occurs around $\mu_S=0.07$ (observed in (a)), indicating that the mutant strain dominates the resident strain in the long run (observed in (b)). }
    \label{fig:R0CovidAndExample}
\end{figure}

To illustrate the theoretical concepts discussed earlier in the context of \gls{covid}, we will utilize the parameter values shown in \cref{tab:CovidPara} 
with the transmission function $\beta=\sech^2(25\mu_S-2)$. \cref{fig:R0CovidAndExample} displays the \gls{r0} for the \gls{covid} model as a function of virulence and host population density for each strain, with $\mu_S=0.1$ and $\mu_S=0.07$. Notably, the mutant strain with virulence $0.07$ outperforms the resident strain with $\mu_S=0.1$, as evidenced by the maximum value of \gls{r0} being attained at around $\mu_S=0.07$.

\begin{figure}[ht]
    \centering
         \begin{subfigure}[b]{0.45\textwidth}
         \centering
         \includegraphics[width=\textwidth]{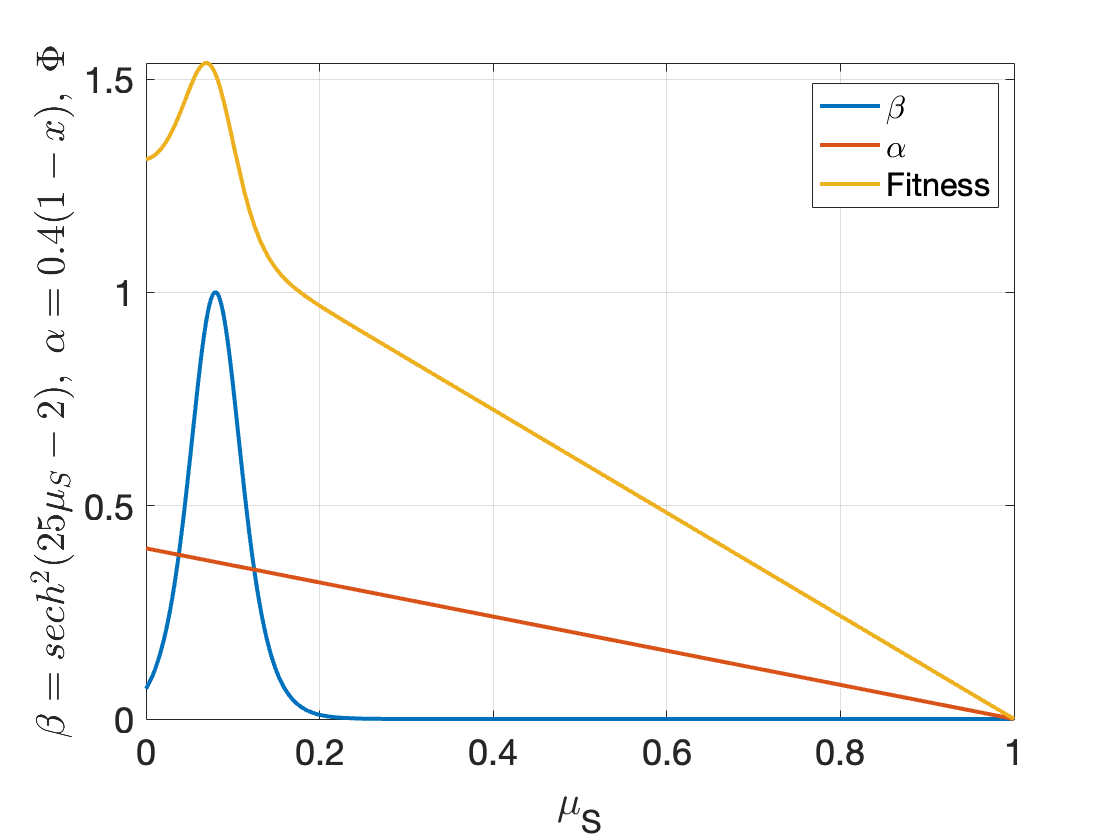}
         \caption{}
     \end{subfigure}
     \hfill
    \begin{subfigure}[b]{0.45\textwidth}
         \centering
         \includegraphics[width=\textwidth]{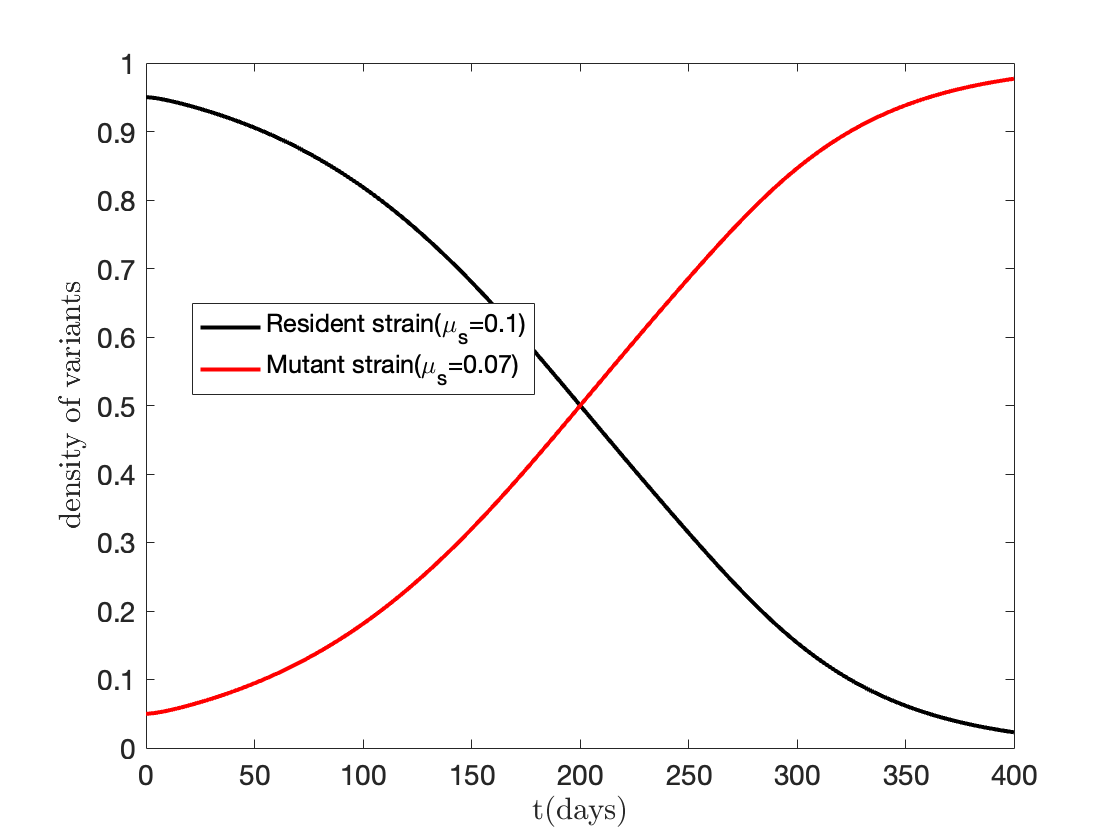}
         \caption{}
     \end{subfigure}

    \caption{We present the results of incorporating the transmission rate $\alpha(\mu_S)$ as a function of virulence in addition to the transmission function $\beta(\mu_S)$, using the same parameter values as in \cref{fig:R0CovidAndExample}. We choose $\beta(\mu_S)=\sech^2(25\mu_S-2)$ and $\alpha(\mu_S)=0.4(1-\mu_S)$. In (a), we show the behavior of the $\Phi=\gls{r0}$ as a function of virulence. In (b), we depict the density of the infected host population for strains with $\mu_S=0.1$ and $\mu_S=0.07$. It is worth noting that, as in the previous case, the mutant strain with virulence $0.07$ dominates the resident strain with $\mu_S=0.1$, as reflected in the maximum of \gls{r0} around $\mu_S=0.07$. In the long run, the density of the mutant strain is higher than that of the resident strain.}
    \label{fig:R0CovidAndExampleAlphaBeta}
\end{figure}

To expand on the concept presented earlier, we consider the possibility of incorporating the transmission rate $\alpha$ as a function of virulence. This allows for a more comprehensive analysis of the evolution of virulence in pathogens. To derive the ESS $\mu_S^*$ in this scenario, we take the derivative of the fitness function $\Phi$ with respect to $\mu_S$. The resulting expression is given by:

\begin{equation}
    \alpha'(\mu_S^*)+\frac{(1-p)\omega}{(v+\mu_S^*)^2} \big(\beta'(\mu_S^*)(v+\mu_S^*) -\beta(\mu_S^*)\big)=0
\end{equation}
where prime ($'$) denotes the derivative of the function with respect to $\mu_S$. \cref{fig:R0CovidAndExampleAlphaBeta} illustrates the fitness function for the virus and the behavior of the infected host density for each strain when $\alpha(\mu_S)=0.4(1-\mu_S)$ and $\beta(\mu_S)=\sech^2(25\mu_S-2)$. Although we have assumed that only $\beta$ and $\alpha$ are affected by virulence for demonstration purposes, readers can extend this analysis to incorporate other factors.

\subsubsection*{Sensitivity analysis}\label{sec:SesAnalys}
First, we will discuss the sensitivity analysis for the \gls{ess} assuming only $\beta$ is a function of $\mu_S$. Hence, the fitness function $\phi$ can be used for this analysis, providing similar results as analyzing $\Phi=\gls{r0}$. We present the analysis using $\phi$ for this case to simplify the calculations. We follow the calculation below to determine the sign of the derivative of \gls{ess} with respect to a parameter $y$ when the fitness function is $\phi$. Note that:
\begin{align*}
    &(v+\mu_S)^2 \frac{\partial \phi}{\partial \mu_S}= C_1(\frac{d \beta}{d\mu_S}(v+\mu_S)-\beta), \text{ and} \\
    &\frac{\partial (v+\mu_S)^2}{\partial y}\big|_{\mu_S=\mu_S^*} \cancelto{0}{\frac{\partial \phi}{\partial \mu_S}\big|_{\mu_S=\mu_S^*}}+ (v+\mu_S)^2 \frac{\partial^2 \phi}{\partial y \partial \mu_S} \big|_{\mu_S=\mu_S^*} = \frac{\partial }{\partial y}\big(C_1(\frac{d \beta}{d\mu_S}(v+\mu_S)-\beta)\big) \big|_{\mu_S=\mu_S^*}.
\end{align*}
Hence, we can conclude the following result (\cref{eq:covidSenRel}) for any parameter $y$ with the \gls{covid} model:
\begin{equation}\label{eq:covidSenRel}
    \frac{d \mu_S^*}{dy} \propto \frac{\partial }{\partial y}\big(C_1(\frac{d \beta}{d\mu_S}(v+\mu_S)-\beta)\big) \big|_{\mu_S=\mu_S^*}.
\end{equation}
Notice that, the $\mu_S^*$ change only with parameter $v$ (if $y=\omega$ or $p$, the right-hand side of \cref{eq:covidSenRel} equals zero at $\mu_S^*$ ), and the ESS level of virulence will always increase as the recovery rate increases, regardless of the exact relationship between the transmission rate $\beta$ and virulence $\mu_S$.

Similarly, if we assume both $\beta$ and $\alpha$ as a function of virulence $\mu_S$,  relationship in the \cref{eq:inva} can be reduced to,
\begin{align}\label{eq:SenAlphaBeta}
    \frac{d\mu_S^*}{d y} \propto \frac{\partial }{\partial y}\big(C_2(\alpha'(v+\mu_S)^2+(1-p)\omega (\beta'(v+\mu_S)-\beta)\big) \big|_{\mu_S=\mu_S^*}
\end{align}
where $C_2=\frac{\epsilon}{(\epsilon+\mu)(\omega+\mu)}$. Hence, the changes in ESS $\mu_S^*$ are unaffected by the changes in transition rate $\epsilon$ (from the exposed group to the infected group) and background death rate $\mu$. When considering the recovery rate $v$, we notice that,
\begin{align*}
    \frac{d \mu_S^*}{d v} \propto  \frac{2 \beta(\mu_S^*)}{(v+\mu_S^*)}-\beta'(\mu_S^*).
\end{align*}
Hence, ESS $\mu_S^*$ will increase with the recovery rate $v$ if and only if $\beta'(\mu_S^*)<\frac{2 \beta(\mu_S^*)}{(v+\mu_S^*)}$.


\subsection*{Reservoir (W.A.I.T.) model type: \glsfmttext{hcv} Example}
\begin{figure}[ht]
    \centering
    \includegraphics[width=\textwidth]{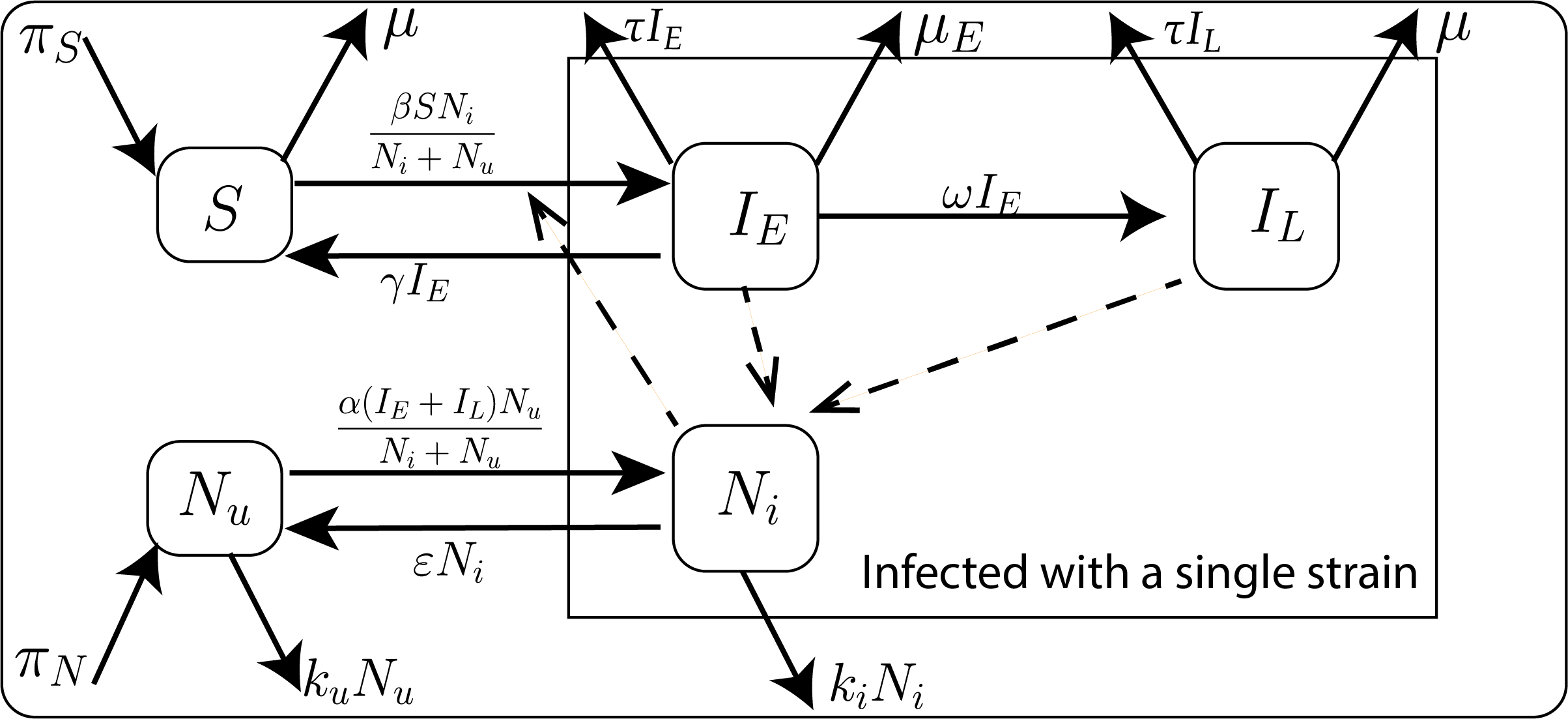}
    \caption{The HCV model is based on a framework for waterborne, abiotic, and indirectly transmitted (W.A.I.T.) disease systems, where an abiotic agent or reservoir (e.g., surface, water supply, or physical instrument) is primary the vector of transmission \cite{miller-dickson_hepatitis_2019, meszaros_direct_2020}. This diagram illustrates the system dynamics of the compartments in relation to \gls{hcv}. Additional details regarding the variables and parameters employed in this diagram are provided in \cref{tab:hcvVara} and  \cref{tab:hcvPara}, respectively. The system of \glspl{ode} that governs this model is presented in \cref{eq:hcvmoel}. }
    \label{fig:HCV}
\end{figure}

We will perform a similar analysis as the \gls{covid} model to the  \gls{hcv} disease dynamics explained by \cref{eq:hcvmoel} and \cref{fig:HCV} (using the single-strain model as explained by \citeauthor{miller-dickson_hepatitis_2019} \cite{miller-dickson_hepatitis_2019}). \cref{eq:hcvmoel} is used to model the dynamics of a specific virus strain in host populations and needle compartments. 

\begin{equation}\label{eq:hcvmoel}
    \begin{aligned}
        \frac{d S}{dt} &= \pi_S+\gamma I_E-\beta S \frac{N_i}{N_i+N_u}-\mu S \\
        \frac{d I_E}{dt} &= \beta S \frac{N_i}{N_i+N_u}-(\omega+\tau+\mu_E+\gamma) I_E \\
        \frac{d I_L}{dt} &= \omega I_E-(\mu+\tau)I_L\\
        \frac{d N_u}{dt} &= \pi_N-\alpha(I_E+I_L)\frac{N_u}{N_i+N_u}-k_uN_u+\epsilon N_i\\
        \frac{d N_i}{dt} &= \alpha(I_E+I_L)\frac{N_u}{N_i+N_u}-k_iN_i-\epsilon N_i \\
    \end{aligned}
\end{equation}
\begin{table}[ht]
\centering
\caption{This table describes variables in \cref{eq:hcvmoel}. All of the variables are measured as the number of people.}
\label{tab:hcvVara}
\begin{tabular}{p{2cm}p{10cm}}
\toprule
\rowcolor{gray!80}
\textbf{Variables}  &\textbf{Description}\\
\midrule
\rowcolor{white}
$S$ & Susceptible individuals who inject drugs and share needles within the community of people who inject drugs (PWID).\\
\rowcolor{gray!30}
$I_E$ & early-stage infected individuals (acute \gls{hcv} infection) \\
\rowcolor{white}
$I_L$ & late-stage-infected individual (chronic \gls{hcv} infection) \\
\rowcolor{gray!30}
$N_u$ & uninfected needles   \\
\rowcolor{white}
$N_i$ & infected needles   \\
\bottomrule
\end{tabular}
\end{table}

\begin{table}[ht]
\centering
\caption{This table describes parameters in \cref{eq:hcvmoel}. The infected death rate is considered as the measure of virulence in the model, and it is highlighted in the table as a row of bold text. }
\label{tab:hcvPara}
\begin{tabular}{p{1.6cm}p{7.3cm}p{2cm}}
\toprule
\rowcolor{gray!80}
\textbf{Parameters} & \textbf{Description} & \textbf{Units}  \\
\midrule
\rowcolor{white}
$\pi_S$ & birthrate of susceptibles  & $\text{person}/\si{day}$  \\
\rowcolor{gray!30}
$\gamma$ & daily fractional self-clearance rate &$\si{day}^{-1}$  \\
\rowcolor{white}
$\mu$ & Natural death rate  & $\si{day}^{-1}$  \\
\rowcolor{gray!30}
$\boldsymbol{\mu_E}$ & \textbf{Virulence (Infected death rate) }&$\boldsymbol{\si{\textbf{day}}^{-1}}$  \\
\rowcolor{white}
$\omega$ & transfer rate into late-stage infection  & $\si{days}^{-1}$  \\
\rowcolor{gray!30}
$\tau$ & rate of entering treatment  & $\si{day}^{-1}$  \\
\rowcolor{white}
$\pi_N$ & birthrate of uninfected needles & $\text{needles}/\si{day}$ \\
\rowcolor{gray!30}
$\epsilon$ & decay rate of \gls{hcv} infection in needles & $\si{days}^{-1}$  \\
\rowcolor{white}
$k_u$ & discard rate of uninfected needles & $\si{day}^{-1}$  \\
\rowcolor{gray!30}
$k_i$ &discard rate of infected needles   & $\si{day}^{-1}$  \\
\rowcolor{white}
$\alpha$ & injection rate times infection of needle probability & $\frac{\text{injections}}{\text{person} .\si{day}}$  \\
\rowcolor{gray!30}
$\beta$ &injection rate times infection of host rate   & $\frac{\text{injections}}{\text{person} .\si{day}}$  \\
\bottomrule
\end{tabular}
\end{table}

Now, consider a two-strain scenario for the \gls{hcv} model, where the strains are denoted as $r$ (resident) and $m$ (mutant). We extend the model by introducing the decision variable $\mu_E$ and assuming that the transmission rate $\beta$ is a function of $\mu_E$, denoted by $\beta_j=\beta(\mu_E^j)$ for $j=r,\ m$. Additionally, we assume that other parameters remain unchanged by evolution. We use the notation $\mathbf{X}_j=(I_{E_j}\ I_{L_j}\ N_{u_j})^T$ to conform with the notation introduced in the theoretical framework. In this context, the matrices $\mathbf{S}\mathbf{F}_j$ and $D_j$ relevant to this problem are given by:
\begin{equation*}
    \begin{aligned}
        \mathbf{S}\mathbf{F}_j=\begin{pmatrix}
            0 &\frac{\alpha S}{N} &\frac{\beta_j S}{N}\\
            0& 0 & 0 \\
            0 &0& 0
        \end{pmatrix}, \  
        \mathbf{D}_j=\begin{pmatrix}
            \epsilon+\mu &0 &0\\
            -\epsilon  & \omega + \mu & 0 \\
            0 &-(1-p)\omega  & v+\mu_S^j
        \end{pmatrix} \ \text{for} \ j=r,m. 
    \end{aligned}
\end{equation*}
We denote the $\bar{S}=\Big(\frac{S}{N_i+N_u}\Big)\Big(\frac{N_u}{N_i+N_u}\Big)$ to simplify the notations carry out in the following calculations.  It should be noted that the basic reproduction number for the single-strain ($r$) model is given by:
\begin{equation}\label{eq:HCVR0}
\rho(\mathbf{S}^*\mathbf{\bar{F}}_r\mathbf{\bar{D}}_r^{-1})=\gls{r0}_r=\sqrt{\frac{\bar{S}^* \beta_r\alpha(\mu_L+\tau+\omega)}{(\omega+\tau+\mu_E^r+\gamma)(k_i+\epsilon)(\mu_L+\tau)}}
\end{equation}
where, $\bar{S}^*$ is the $\bar{S}$  at the \gls{dfe}. Consider the resident strain at the mutant-free equilibrium $\hat{V}_r$. It can be observed that $|\mathbf{S}\mathbf{F}_r-\mathbf{D}r|_{\hat{V}_r}=0$, indicating that the following expression holds true:
\begin{equation}\label{eq:HCVSusMFE}
\hat{\overline{S}}_r=\bar{S}|_{\hat{V}_r}=\frac{(\omega+\tau+\mu_E^r+\gamma)(k_i+\epsilon)(\mu_L+\tau)}{ \beta_r\alpha(\mu_L+\tau+\omega)}
\end{equation}
Therefore, the basic reproduction number for the MFE of
the resident strain can be derived using the following equation:
\begin{equation}\label{eq:HCVR0MFE}
\rho(\hat{\mathbf{S}}_r\mathbf{\bar{F}}_m\mathbf{\bar{D}}_m^{-1})=\gls{r0}(\hat{V}_2)=\sqrt{\frac{\hat{\overline{S}}\beta_m\alpha(\mu_L+\tau+\omega)}{(\omega+\tau+\mu_E^m+\gamma)(k_i+\epsilon)(\mu_L+\tau)}}
\end{equation}
By substituting \cref{eq:HCVSusMFE} into the expression for the basic reproduction number at \gls{mfe} (\cref{eq:HCVR0MFE}), we obtain a form:
\begin{equation} \label{eq:covidR0MFE2}
\gls{r0}(\hat{V}_2)=\frac{\Phi_m}{\Phi_r}
\end{equation}
where $\Phi_j=\gls{r0}_j$. This means that the available basic reproduction number information, modified with the parameters specific to evolution, may able to directly used to determine the \gls{ess} in more complex models with reservoir. In the event where only $\mu_E$ and $\beta$ parameters are subject to evolution, the expression for $\gls{r0}(\hat{V}_2)$ can be simplified to the form $\gls{r0}(\hat{V}_2)=\sqrt{\frac{\phi(\mu_E^m)}{\phi(\mu_E^r)}}$ where the function $\phi(\mu_E)=\frac{\beta(\mu_E)}{\omega+\tau+\mu_E+\gamma}$. In this example, we assume $\mu_E$ is the decision variable, and $\beta$ is a function of $\mu_E$. Since $\argmax \sqrt{\phi(x)}=\argmax \phi(x)$, ESS can be obtained by analyzing $\phi(\mu_E)$ and we will consider it as the fitness function for the strain. Hence, ESS, denoted $\mu_E^*$, is given by,
\begin{equation}
    \beta'(\mu_E^*)=\frac{\beta(\mu_E^*)}{\omega+\tau+\mu_E+\gamma},
\end{equation}
where prime ($'$) denote the derivative with respect to $\mu_E$. 
\subsubsection*{Sensitivity analysis}
In a similar calculation to the \gls{covid} model example, a derivative of ESS with respect to a given parameter $y$ can be explained by,
\begin{equation}\label{eq:HCVSEn}
    \frac{d \mu_E^*}{d y}  \propto \frac{\partial }{\partial y}\big(\beta'(\omega+\tau+\mu_E+\gamma)-\beta\big) \big|_{\mu_E=\mu_E^*}.
\end{equation}
Hence, ESS is only sensitive to the parameters $\omega,\ \tau, \gamma$, and $\beta$. Furthermore, it proves that $\frac{d \mu_E^*}{d y}>0$ with parameters  $y=\omega,\ \tau$ or  $\gamma$. Therefore, we can conclude that the ESS level of virulence will increase as;
\begin{itemize}
\item The treatment rate $\tau$,
\item The transfer rate into late-stage infection $\omega$, or
\item The self-clearance rate $\gamma$ 
\end{itemize}
increases. In addition, when $y=\beta$, the \cref{eq:HCVSEn} can be reduced to,
\begin{equation}
    \frac{d \mu_E^*}{d \beta} \propto \big( (\omega+\tau+\mu_E+\gamma)\frac{\beta''(\mu_E^*)}{\beta'(\mu_E^*)}-1)\big)<0
\end{equation}
where, ($''$) denotes the second derivative with respect to the $\mu_E$. Since the maximum of the fitness function has been attained at $\mu_E^*$, the second derivative condition can be reduced to $\beta''(\mu_E^*)<0$. Therefore, $\frac{d \mu_E^*}{d \beta}<0$, and the ESS level of virulence will decrease as the infection of host rate $\beta$ increases.

\section*{Discussion}
In this study, we utilize mathematical approaches to identify the evolutionary stable strategy (ESS) level of virulence for virus pathogens of differing structures and natural history: SARS-CoV-2 and HCV. The viruses underlying these outbreaks exhibit distinct characteristics in the diseases they cause, their modes of transmission, biological structures, and the level of virulence exerted on their hosts. We utilize mathematical models of each, with parameters determined from existing models, that implement published data \cite{ogbunugafor2020variation, miller-dickson_hepatitis_2019}. Furthermore, we  propose a framework for identifying evolutionary stable strategies based on the identification of a fitness function that aids in the invasion analysis of mutant strains. We apply the invasion analysis developed in other texts \cite{otto2011biologist} to compute ESS virulence in different viral natural histories. 

In SARS-CoV-2, we examine two different cases:(i) one in which virulence is a function solely of the $\beta$ transmission parameter, where symptomatic individuals transmit to susceptible individuals, and (ii) another where virulence is a function of terms associated with both symptomatic ($\beta$) and asymptomatic ($\alpha$) individuals (See \cref{eq:covidSenRel,eq:SenAlphaBeta}).
This duality recapitulates debates early in the COVID-19 pandemic, where experts sought to identify the role of asymptomatic infection in disease dynamics \cite{moghadas2020implications, mizumoto2020estimating, nishiura2020estimation,kronbichler_asymptomatic_2020,rothe_transmission_2020}. Our findings highlight why properly characterizing the transmission mechanism of an emerging infectious disease is so crucial: they offer profoundly different predictions for how ESS virulence will evolve.

When virulence is a function of transmission from symptomatic individuals, the ESS level of virulence increases as a function of recovery rate. In this grim hypothetical scenario, virulence goes up as recovery rate goes up, suggesting that treatments and public health interventions will foster increased virulence. Encouragingly (from a public health perspective), SARS-COV-2 is now widely understood to have more complicated transmission dynamics, with both symptomatic and asymptomatic transmission playing a role \cite{mizumoto2020estimating, nishiura2020estimation,kronbichler_asymptomatic_2020}. Here, our observations for ESS virulence are more complicated, with the direction of virulence evolution depending on several other mathematical relationships. For example, ESS virulence depends on the slope of  the transmission rate satisfying a very particular set of conditions, including the asymptomatic recovery rate.  These findings in SARS-CoV-2 highlight the complexity of phenotypic evolution in emerging pathogens: simplistic or narrow views of SARS-CoV-2 natural history are likely inadequate, miss the many nuances and dependencies that define how a given population of viruses will evolve in a population of hosts, and implore much more careful definitions and examinations of virulence. 

We also compute the ESS virulence for HCV transmission in a population of persons who inject drugs (PWID). The model of HCV transmission includes an indirect transmission dimension, where infected needles circulate in a population. Using an existing model of HCV dynamics, we observe that the ESS virulence will increase as the treatment rate, rate of movement into the late-stage infection increases, and the self-clearance rate increases. Despite being a public health concern, HCV has highly effective treatments available on the market \cite{liang2013current,morozov_hepatitis_2018,hu_hepatitis_2020}. Intriguingly, self-clearance rate of HCV is known to be influenced by host genetics, with alleles fostering increased or impaired rates of clearance \cite{thomas2009genetic, prokunina2013variant, ge2009genetic}. Our findings highlight why one might expect that virulence evolution may increase in populations of individuals who carry the high-clearance allele under certain conditions. This is an intriguing finding because of what it says about how heterogeneity in host characteristics may influence the trajectory of virus evolution. This finding has been discussed in other settings \cite{hebert2020beyond}. Further, the HCV model highlights how diseases with an indirect transmission route differ from direct pathogens. Future studies in this realm will examine ESS virulence evolution in epidemics with other structures–waterborne, vector-borne, and classical fomite transmission.

This study is not designed to explain any particular outcome or directly inform public health interventions. However, we do acknowledge that the use of well-known human pathogens as examples comes with the risk of misinterpretation. Therefore, we emphasize (rather strongly) that this study only aims to utilize models  with good existing data, to make a general point about the capriciousness of virulence evolution, as it strongly depends on several features of the virus's natural history and particulars of transmission. We feel that this is an important point to make in light of misinterpretations of modern findings in infectious diseases, some of which even fuel misinformation. 

Moreover, this point transcends any particular virus-host system: we believe that attempts to understand any pathogen-host system will be colored by similar nuances. Furthermore, this study utilizes analytical approaches to study disease dynamics, which are increasingly understood to be defined by nuances that undermine the assumptions of SEIR-style analytical descriptions. While this problem plagues many areas of infectious disease modeling, we acknowledge that it undermines the realism of these simulations. Nonetheless,  mathematical modeling remains an important tool for studying disease dynamics because it provides a transparent means to engage the actors that drive infectious outbreaks. Future investigations can utilize other computational approaches for modeling infectious diseases \cite{marshall_formalizing_2015, cardenas_genomic_2022}, and examine a growing number of virus (and other pathogen) evolution scenarios. 

\section*{Conclusion}
Our study hopes to add to a growing chorus to refine aspects of the evolution of virulence, a canon in evolutionary theory that has spawned an entire subfield at the intersection of evolutionary biology and epidemiology. While it has helped to revolutionize our understanding of infectious diseases by offering an evolutionary lens on the host-pathogen interaction, it can sometimes oversimplify how pathogen evolution manifests. Our study focuses on outbreaks caused by single-stranded RNA viruses and still offers diverse patterns in evolutionary outcomes. In response to these (and other) findings, those interested in pathogen evolution should more carefully consider their definition of virulence, what aspects of the disease's natural history underlie it, and how it may evolve in a given setting. Even more, our findings are consistent with discussions in the broader fields of ecology where experts continue to examine the meaning and consequences of context-dependence \cite{catford_addressing_2022}. 

\section*{Acknowledgements}
The authors would like to acknowledge members of the Ogbunu Lab, S. Almagro-Moreno, S. Scarpino, L. Zaman, and W. Turner for their helpful input on the topic. SS acknowledges support from the Seesel Postdoctoral Fellowship from Yale University. CBO acknowledges support from the MLK Visiting Scholars and Professors Program from the Massachusetts Institute of Technology. 

\section*{Author Contributions}
Project conception: S.S., K.K. and C.B.O. Model development: S.S., C.B.O. Analysis and Interpretation: S.S., K.K. and C.B.O. Supervision: P.E.T., C.B.O. Writing: S.S., K.K., P.E.T., C.B.O.

\printunsrtglossary[type=abbreviations]

\printunsrtglossary[type=symbols]

\printbibliography


\end{document}